\documentclass[twocolumn]{aastex63}

\usepackage{amsmath}
\usepackage{mathtools}
\usepackage{tabularx}
\usepackage[hyphens]{url}
\usepackage{graphicx}
\usepackage{subfigure}
\usepackage[figuresright]{rotating}
\usepackage{booktabs}
\usepackage{multirow}
\newcommand{\erfc}[1]{\operatorname{erfc}\left(#1\right)}

\received{XXX}
\revised{XXX}
\accepted{XXX}

\submitjournal{ApJ}

\shorttitle{MaDCoWS2 DR1}
\shortauthors{Thongkham et al.}

\graphicspath{{./}{figures/}}

\begin{document}

\title{The Massive and Distant Clusters of WISE Survey 2: Equatorial First Data Release}

\author[0000-0001-7027-2202]{Khunanon Thongkham}
\affiliation{Department of Astronomy, University of Florida, 211 Bryant Space Center, Gainesville, FL 32611, USA}

\author[0000-0002-0933-8601]{Anthony H. Gonzalez}
\affiliation{Department of Astronomy, University of Florida, 211 Bryant Space Center, Gainesville, FL 32611, USA}
 
\author[0000-0002-4208-798X]{Mark Brodwin}
\affiliation{Department of Physics and Astronomy, University of Missouri, 5110 Rockhill Road, Kansas City, MO 64110, USA}
 
\author[0000-0003-3428-1106]{Ariane Trudeau}
\affiliation{Department of Astronomy, University of Florida, 211 Bryant Space Center, Gainesville, FL 32611, USA}
 
\author{Ripon Saha}
\affiliation{Department of Physics and Astronomy, University of Missouri, 5110 Rockhill Road, Kansas City, MO 64110, USA}
\affiliation{Wells Fargo, 550 S Tryon St, Charlotte, NC 28213, USA}

\author{Peter Eisenhardt}
\affiliation{Jet Propulsion Laboratory, California Institute of Technology, 4800 Oak Grove Dr., Pasadena, CA 91109, USA}

\author[0000-0003-0122-0841]{S. A.\ Stanford}
\affiliation{Department of Physics, University of California, One Shields Avenue, Davis, CA 95616, USA}

\author[0000-0001-9793-5416]{Emily Moravec}
\affiliation{Green Bank Observatory, P.O. Box 2, Green Bank, WV 24944, USA}

\author[0000-0002-7898-7664]{Thomas Connor}
\affiliation{Jet Propulsion Laboratory, California Institute of Technology, 4800 Oak Grove Dr., Pasadena, CA 91109, USA}
\affiliation{Center for Astrophysics $\vert$\ Harvard\ \&\ Smithsonian, 60 Garden St., Cambridge, MA 02138, USA}

\author[0000-0003-2686-9241]{Daniel Stern}
\affiliation{Jet Propulsion Laboratory, California Institute of Technology, 4800 Oak Grove Dr., Pasadena, CA 91109, USA}

\begin{abstract}
The Massive and Distant Clusters of \textit{WISE} Survey 2 (MaDCoWS2) is a new survey designed as the successor of the original MaDCoWS survey. MaDCoWS2 improves upon its predecessor by using deeper optical and infrared data and a more powerful detection algorithm (PZWav). As input to the search, we use $grz$ photometry from DECaLS in combination with W1 and W2 photometry from the CatWISE2020 catalog to derive the photometric redshifts with full redshift probability distribution functions for \textit{WISE}-selected galaxies. Cluster candidates are then detected using the PZWav algorithm to find three-dimensional galaxy overdensities from the sky positions and photometric redshifts. This paper provides the first MaDCoWS2 data release, covering $1461$ ($1838$ without masking) deg$^2$ centered on the Hyper-SuprimeCam Subaru Strategic Program equatorial fields. Within this region, we derive a catalog of $22,970$ galaxy cluster candidates detected at S/N $>5$. These clusters span the redshift range $0.1<z<2$, including $1312$ candidates at $z>1.5$. We compare MaDCoWS2 to six existing catalogs in the area. We rediscover $60\%-92\%$ of the clusters in these surveys at S/N $>5$. The medians of the absolute redshift offset are $<0.02$ relative to these surveys, while the standard deviations are less than $0.06$. The median offsets between the detection position from MaDCoWS2 and other surveys are less than $0.25$ Mpc. We quantify the relation between S/N and gas mass, total mass, luminosity, and richness from other surveys using a redshift-dependent power law relation. We find that the S/N-richness relation exhibits the lowest scatter. 
\end{abstract}


\section{Introduction} \label{sec:Introduction}
Galaxy clusters are one of the most important tools to constrain cosmological parameters \citep{2009Vikhlinin,2011Allen}, and to understand galaxy evolution \citep{2010von} and feedback processes \citep{2012Fabian,2012McNamara}. Large statistical samples of galaxy clusters are the necessary building blocks for such studies. From the early days of large optical cluster surveys by \cite{1958Abell,1961Zwicky,1989Abell}, cluster searches have grown into multi-wavelength endeavors spanning a large range of redshift. Through the overdensity of galaxies, a great number of optical/infrared detected clusters have been produced by the Sloan Digital Sky Survey \citep[SDSS;][]{2007Koester,2009Wen,2012Wen,2014Oguri,2014Rykoff,2018Banerjee}, and \textit{WISE} \citep{2018Wen,2019Gonzalez}. Using the X-ray emission for the intracluster medium (ICM), the \textit{ROentgen SATellite (ROSAT)} all-sky X-ray survey has provided a large sample of clusters \citep{199Ebeling,2001Ebeling,2013Bohringer,2016Pacaud,2017Bohringer}. The \textit{Planck} mission \citep{2016Planck}, the South Pole Telescope \citep[SPT;][]{2015Bleem,2020Bleem,2019Bocquet,2020Huang,2023Klein}, and the Atacama Cosmology Telescope \citep[ACT;][]{2018Hilton,2021Hilton} have also delivered large cluster catalogs found by the Sunyaev–Zel’dovich (SZ) effect at millimeter wavelengths. The primary advantage of optical/infrared searches is that they are able to detect lower mass clusters and groups at high redshift than SZ surveys, while probing a large comoving volume \citep{2016Pacaud,2019Gonzalez,2020Dicker,2021Hilton}. A wide optical/infrared survey with deep photometry has the potential to provide a complete catalog of galaxy clusters over a wide range of cluster mass and redshift.

 The redshift range $z$ $\sim 1.5-2$ is the era when the transition from the protoclusters to galaxy clusters occurs for massive clusters \citep{2016Overzier}.  Cosmological simulations show that the collapse of the most massive protoclusters starts at $z$ $ \sim 1.5-2$ \citep{2013Chiang,2017Chiang,2020Rennehan}. A number of studies imply that this period is when environmental quenching becomes important in clusters. Multiple authors \citep[e.g.,][]{2010Tran,2010Hilton,2013Brodwin,2014Alberts,2016Alberts} have shown that star formation in the cluster core galaxies gradually increases at $z$ $>1$ relative to the field. Compared to low redshift galaxy clusters which are filled with quenched galaxies \citep{1969Rood,1987Binggeli}, galaxy clusters at $z$ $>1$ have lower quenched galaxy fractions \citep{2017Nantais}. Also, star formation in galaxies at cluster cores quenches around $z$ $\sim 1.5-2$ \citep{2009Mei,2010Rettura,2012Snyder,2014Newman}.  However, the current sample size is only in the hundreds. Larger samples of galaxy clusters at $z$ $\sim 1.5-2$ are needed to further our understanding of the evolution of the galaxies, gas, and dark matter in galaxy clusters during this epoch of formation. Combining the sample in this range with samples at lower redshift enables one to trace the full history of galaxy cluster assembly.

Surveys by \textit{Spitzer} \citep{2010Papovich,2012Stanford,2012Zeimann,2013Muzzin,2015Webb,2016Cooke}, the \textit{X-ray Multi-Mirror Mission} \citep[\textit{XMM-Newton};][]{2014Andreon,2014Mantz}, ACT \citep{2018Hilton,2021Hilton}, and SPT \citep{2019Strazzullo,2019Bocquet,2023Klein} have provided galaxy clusters at $z$ $>1.5$. The sample size at this range however is still far smaller than the sample size at $z$ $<1.5$ for all these surveys. Searches with \textit{Spitzer}, which are limited to $10$ deg$^2$, are available in the \textit{Spitzer} Deep, Wide-field Survey \citep[SDWFS;][]{2009Ashby}. \cite{2021Wen} provide a catalog of galaxy cluster candidates in this range using data from the Subaru Hyper Suprime-Cam (HSC) and unWISE-produced using photometric redshifts. The catalog contains $642$ cluster candidates at $z$ $>1.5$. Their work highlights the potential of deep infrared and optical data in cataloging galaxy clusters at high redshift. To provide a larger sample of galaxy clusters extending to high redshift, a larger area is needed. Using the DESI Legacy Imaging Survey \citep[LS;][]{2019Dey} instead of the Subaru Hyper Suprime-Cam data enables a wider search area.

We introduce a cluster search program called the Massive and Distant Clusters of \textit{WISE} Survey 2, or MaDCoWS2. MaDCoWS2 uses infrared data from CatWISE2020 \citep{2021Marocco} in conjunction with optical data from LS. The search continues the legacy of MaDCoWS \citep{2019Gonzalez}, which detects clusters at $ 0.7 \;\leq z \leq\;1.5$ over the full extragalactic sky. The MaDCoWS catalog enabled study of the stellar mass fractions in clusters \citep{2019Decker,2021Decker}, the impact of environment on active galactic nuclei (AGN) \citep{2018Mo,2019Moravec,2020Moravec_a,2020Moravec_b,2020Mo}, and calibration of the galaxy cluster mass to richness relations \citep{2020Dicker}. The success of MaDCoWS relies on AllWISE 2013 W1 and W2 which have $5\sigma$ sensitivities of $16.63$ and $15.47$ (Vega), and Pan-STARRS$3\pi$ $g$, $r$, and $z$ which have $5\sigma$ sensitivities of $23.3$, $23.2$, and $22.3$. It also depends on color and magnitude cuts to isolate $z$ $>0.7$ galaxies. MaDCoWS2 aims to extend MaDCoWS success and discover clusters from $z$ $=0.1$ to $2$ using deeper photometry and improved photometric redshifts. CatWISE2020 provides W1 and W2 sensitivity of $17.43$ and $16.47$ ($5\sigma$). LS provides $g$, $r$, and $z$ sensitivity of $24.9$, $24.2$, and $23.3$ ($5\sigma$). Moreover, MaDCoWS2 has better performance than MaDCoWS because it finds clusters using photometric redshifts instead of color and magnitude cuts. The algorithm used to detect clusters in MaDCoWS2 is one of the algorithms that will be used in the \textit{Euclid} mission \citep{2019Euclid} galaxy cluster survey. The main advantage of MaDCoWS2 is its reach in redshift ($z$ $>1.5$) and the comoving volume it explores. 

In this paper, we present the first data release from MaDCoWS2, spanning $\sim$ 1500 deg$^2$. The final survey will cover $>8000$ deg$^2$. We focus on the wide areas overlapped with the fall and spring equatorial fields of the Hyper Suprime-Cam Subaru Strategic Program Wide survey. The structure of this paper is as follows. In section \ref{sec:data}, we give an overview of the data used in the survey. Section \ref{sec:data prep} explains the method we use to prepare our photometric data for the photometric redshift generation. In section \ref{sec:photo-z}, a brief description of the photometric redshifts is provided. In section \ref{sec:cluster finding}, we discuss the cluster-finding algorithm and how we deal with contamination in our search. Section \ref{sec:Basic characteristics} discusses the basic properties of our cluster catalog, while section \ref{sec:survey characterization} characterizes our catalog relative to other cluster catalogs as well as a simulation.  Lastly, we summarize the key results of this paper in section \ref{sec:summary}.

The cosmology used throughout this paper follows the $\Lambda$CDM cosmology from \cite{2020Planck} with $\Omega_m =0.315$ and $H_0 = 67.4$ km s$^{-1}$ Mpc$^{-1}$. Magnitudes are in the Vega system unless stated otherwise.

\section{Data} \label{sec:data}
For MaDCoWS2, we use the CatWISE2020 infrared catalog \citep{2020Eisenhardt,2021Marocco} and the DESI LS optical catalog \citep{2019Dey} as the basis for the input galaxy catalog and the photometric redshifts obtained for the search. We also use the deeper optical data from Hyper Suprime-Cam (HSC) Subaru Strategic Program as a tool for validation. In this section, we give an overview of the data sets used in this survey.  

\subsection{CatWISE} \label{subsec:catwise}

CatWISE2020 \citep{2020Eisenhardt,2021Marocco} is a program to construct an all-sky catalog of sources in W1 ($3.4 \mu \text{m}$) and W2 ($4.6 \mu \text{m}$) from the \textit{Wide-field Infrared Survey Explorer} \cite[\textit{WISE};][]{2010Wright} and \textit{Near-Earth Object Wide-field Infrared Survey Explorer Reactivation Mission} \cite[\textit{NEOWISE};][]{2014Mainzer} survey data. \cite{2020Eisenhardt} gives a full overview of the preliminary catalog, which is based on 8 sky coverages with WISE, while \cite{2021Marocco} provides the official CatWISE2020 catalog, which is available via the IRSA database at the NASA Infrared Processing and Analysis Center $\left(\text{IPAC} \right)$.\footnote{\url{https://irsa.ipac.caltech.edu/cgi-bin/Gator/nph-scan?submit=Select&projshort=WISE}}

The original MaDCoWS survey used the AllWISE catalog \citep{2013Cutri}. 
 The CatWISE2020 photometry comes from unWISE \citep{2014Lang} coadds of 12 sky coverages from the AllWISE and NEOWISE 2019 data release.\footnote{\url{https://wise2.ipac.caltech.edu/docs/release/neowise/neowise_2019_release_intro.html}} 
  The details of the construction of unWISE coadds are presented in \cite{2019Meisner}. As described in \cite{2021Marocco}, initial detection positions for CatWISE2020 are based upon the \texttt{crowdsource} detection algorithm \citep{2018Schlafly} which was also used by the unWISE catalog \citep{2019Schlafly}. CatWISE2020 merges the \texttt{crowdsource} W1 and W2 detection lists into a single list, treating sources as identical if they are within 2.4", and then simultaneously solves for the source position and W1 and W2 fluxes using point spread function fitting \citep{2020Eisenhardt}. The sources in the catalog have $\text{S/N} \geq 5$ in either W1 or W2 and contain no identified artifacts. The catalog comprises approximately $1.89$ billion sources covering the entire sky. The $5\sigma$ %S/N 
 photometric limits of the CatWISE2020 catalog are W1 $=17.43$ and W2 $=16.47$, while the catalog $90\%$ completeness limits are W1 $=17.7$ and W2 $=17.5$ \citep{2021Marocco}, based on comparison with the \textit{Spitzer} South Pole Telescope Deep Field survey \citep{2013Ashby}.

The \texttt{crowdsource} algorithm  is designed to detect point sources. Thus, sources resolved at \textit{WISE} resolution may be spuriously deblended, resulting in numerous fragmented detections for nearby galaxies. These fragmented detections typically do not have optical counterparts and can be mistaken as high-redshift galaxies. In \S \ref{sec:Fragmented Galaxies}, we discuss the mitigation of this effect.

\subsection{DESI Legacy Imaging Surveys} \label{sec:legacy survey}
The optical data used in this search are the $g$, $r$, and $z$ bands from the DECam Legacy Survey (DECaLS) which is a part of the DESI Legacy Imaging Surveys, (LS). A complete description of these surveys is given in \cite{2019Dey}. The project is a combination of data from DECaLS, the Beijing-Arizona Sky Survey (BASS), and the Mayall $z$-band Legacy Survey (MzLS). The total coverage area of the surveys together is $14,000\;\text{deg}^2$ for $|b| > 18^\circ$ in Galactic coordinates and $-18^\circ < \delta < +84^\circ$ in celestial coordinates. DECaLS covers more than $10,000\;\text{deg}^2$ at $ \delta < 32.375^{\circ}$. Hereafter, LS denotes the DECaLS segments of the DESI Legacy Imaging Surveys.

We use the DR9 catalog,\footnote{\url{https://www.legacysurvey.org/dr9/description/}} which provides not only source positions and photometry, but also morphological classification for a significant number of sources. For each source, parameters related to the shape model fitted to the source are included. These features become useful in the cluster search since extended nearby galaxies are often fragmented into multiple detections in the infrared catalog. 

\subsection{Hyper Suprime-Cam Subaru Strategic Program} \label{subsec:HSC}
For this initial data release, we have focused on a region that includes the footprint of the equatorial fields of the HSC Subaru Strategic Program \citep{2018Aihara,2018Aiharab}. The deeper optical data in the HSC region are valuable for validation of our cluster catalog, and the HSC CAMIRA catalog \citep{2018Oguri} provides a comparison cluster catalog based upon this deeper optical data. The HSC public data release 3 \citep{2022Aihara} covers $1,470\;\text{deg}^2$, with approximately $1300$ deg$^2$ within the equatorial fields. The HSC Wide catalog provides $g$, $r$, $i$, $z$, and $y$ bands photometry with $5\sigma$ limits of $26.5$, $26.5$, $26.2$, $25.2$, and $24.4$ (AB). For the HSC Deep/UltraDeep catalog, the corresponding  $5\sigma$ limits are $27.4$, $27.1$, $26.9$, $26.3$, and $25.3$ (AB). 

\section{Matching CatWISE2020 and LS Catalogs} \label{sec:data prep}
To generate the photometric redshifts, we need to create a joint photometric catalog from CatWISE2020 and LS.  We also need to filter out spurious sources to minimize potential contamination in the final cluster catalog.

\subsection{Catalog Matching} \label{subsec:Catalog Matching}
\begin{figure}[htbp!]
\centering
\includegraphics[width=0.9\columnwidth]{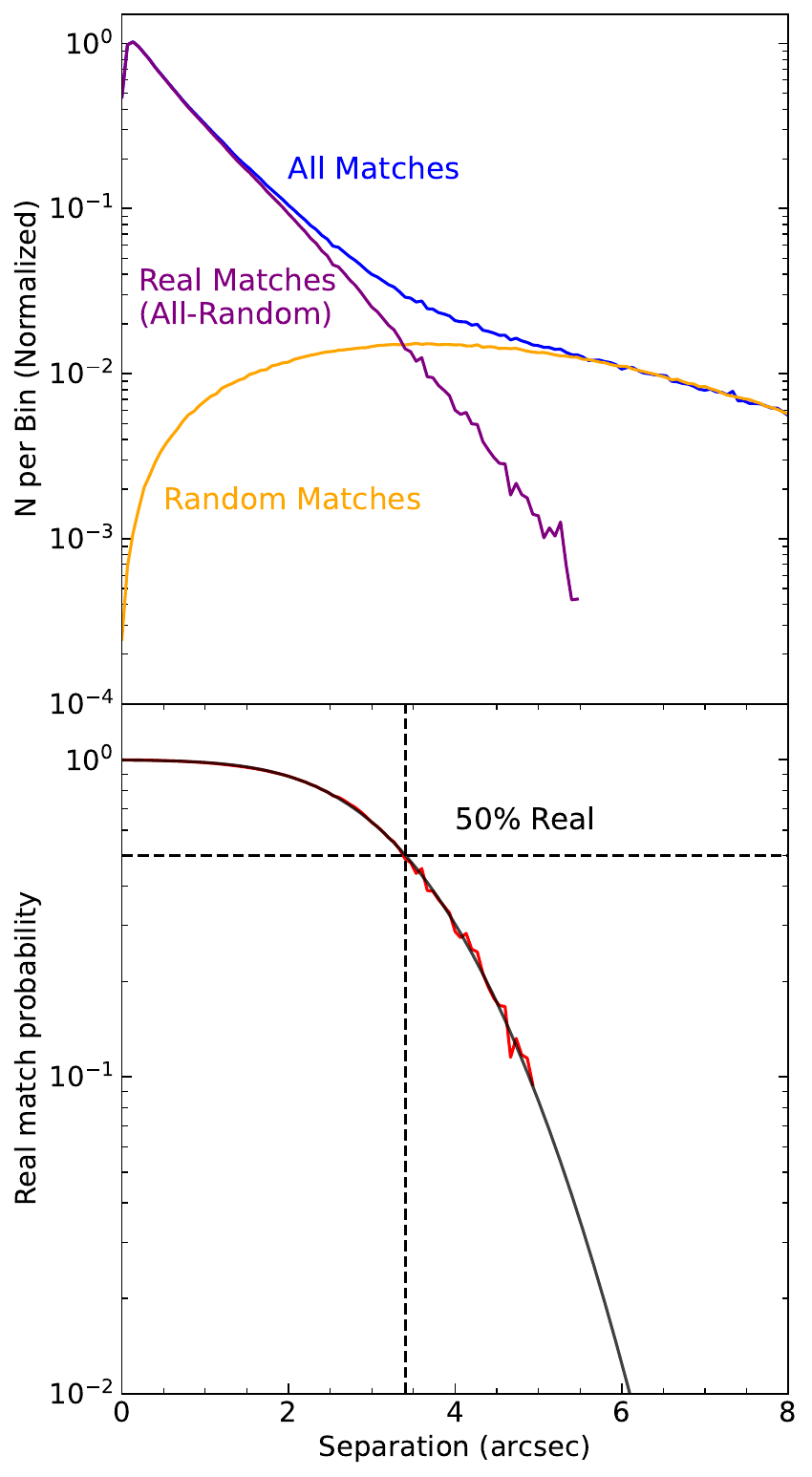}
\caption{The distribution of the distance of the nearest match generated for a tile centered at $\alpha=15^\circ$ and $\delta=-35^\circ$. Top: The distribution of the match separation ($s$) of our actual data is in blue. The orange curve is a random match generated by shifting the CatWISE2020 catalog $5$ arcminutes in Right Ascension and cross-matching with the original catalog. The purple curve, which we denote as real matches, is the difference in number between the actual match and the random match. Separations are put into bins of $0.067$ arcseconds. Both the blue and purple curves are normalized to an integral of $1$ over the range of the histogram, $3\times10^{-4}<s(^{\prime\prime})<10$. Assuming most of the matches at large $s$ are random, the orange plot is normalized to the blue plot at large $s$, $5.75<s(^{\prime\prime})<6.5$. The normalization ensures that the purple curve represents real matches. Since blue and orange curves consist of the same number of galaxies, the blue curve without normalization would be lower at large radii as it has more galaxies at lower radii.} Bottom: The probability that a match is real, as defined in Equation \ref{eq:frac0}, is shown in red. We stop plotting the red curve when the probability of a real match is $< 0.1$. The black curve is an error function that is fit to the data, demonstrating that the probability that a match is real can be well-modeled with this function. 
\label{fig:sep}
\end{figure}

We use CatWISE2020 as the reference catalog and match LS photometry to CatWISE2020 sources. We use the nearest positional match to CatWISE2020 from LS for all sources via the \texttt{astropy~coordinates} package. In case of multiple matches (i.e., the same LS source matched to multiple CatWISE2020 objects), we only use the closest match to our reference. The catalog matching is performed in tiles. We construct these matched photometric catalog tiles by dividing the sky into $10^\circ \times 10^\circ$ tiles of equal RA and DEC. The blue curve in Figure \ref{fig:sep} (top panel) displays the distribution of the distance of the nearest match ($s$) for galaxies in one of our survey tiles centered at $\alpha=15^\circ$ and $\delta=-35^\circ $. We expect real associations to dominate these matches at the smallest $s$ while spurious matches dominate these matches at the largest $s$. The orange curve in Figure \ref{fig:sep} is the same distribution, but for cross-matching the galaxy catalog to a random catalog created by shifting CatWISE2020 sources by $5^{\prime}$ in Right Ascension. Assuming both curves consist of mainly random matches at large $s$, we normalize the random distribution in the orange curve to match the observed (blue) distribution at  $s=5.75^{\prime\prime}-6^{\prime\prime}$. The difference between these two curves is the distribution of $s$ for real matches (purple). In the bottom panel, the fraction of real matches, calculated as
\begin{equation}
    f_{real} = 1-\frac{r}{t},   
\label{eq:frac0}
\end{equation}
is shown as the red curve where $r$ is the number of random matches and $t$ is the number of real matches. We fit a complementary error function (black) to the red curve. The function is described by 
\begin{equation}
    \erfc{z} = 1-\frac{2}{\sqrt{\pi}}\int_0^{z}e^{-t^2}dt \;\text{, where}\; z=\frac{s-\mu}{\sigma}    
\end{equation}
 with $\mu$ and $\sigma$ being the mean and standard deviation of a Gaussian distribution. The best fit for the red curve is for $\sigma=3.40\pm0.01^{\prime\prime}$ and $\mu=1.64\pm0.02^{\prime\prime}$. The quality of the fit indicates that a Gaussian error distribution can be used to model the probability that a match between CatWISE2020 and Legacy Survey sources is real. 

\begin{figure}[htbp]
\centering
\includegraphics[width=0.9\columnwidth]{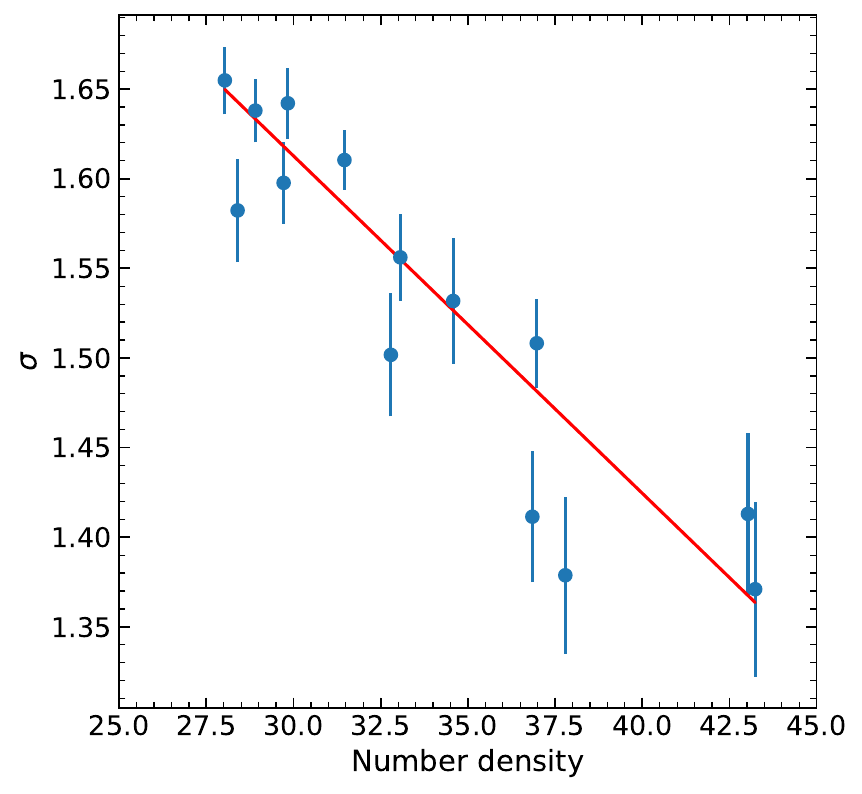}
\caption{Width, $\sigma$, of the complementary error function fitted to the probability that a cross-match is real as a function of source number density. The number densities are in units of $10^3$ deg$^{-2}$.}
\label{fig:rad_trend}
\end{figure}

We note that $\sigma$ and $\mu$ change with the density of sources in the sky because spurious matches become more probable in crowded fields. Figure \ref{fig:rad_trend} illustrates the variation of $\sigma$, the width of the error function, with source density. The red line in Figure \ref{fig:rad_trend} is a linear fit for the dependence of $\sigma$ on source density. The fit values are used to obtain $\sigma$ and $\mu$ for other tiles in this survey. This allows us to account for the change in $\sigma$ and $\mu$ with source density. 
The error functions help us quantify the effect of the potential mismatch on the probability distribution function (PDF) of a source redshift later in \S \ref{sec:photo-z}. 

    \subsection{Catalog Filtering}
\label{sec:filtering}
Before we input matched catalogs into the search algorithm, we remove objects in the catalog that may be spurious detections.
Diffraction spikes, scattered-light halos, optical ghosts, and charge persistence from bright stars can contaminate \textit{WISE} photometry. We employ \texttt{cc}$\_$\texttt{flag} and \texttt{ab}$\_$\texttt{flag} from CatWISE2020 to remove likely contaminants from the photometric catalog. We allow only objects with \texttt{cc}$\_$\texttt{flag} and \texttt{ab}$\_$\texttt{flag} value equal to $0$ (no artifact).  

Extended galaxies cause spurious detection, leading to incorrect estimates of high photometric redshifts (see \S \ref{sec:Fragmented Galaxies} for more explanation). While we already mitigate this by applying galaxy masks on our density maps as well as rejecting spurious candidates (more detail in \S \ref{sec:Fragmented Galaxies}), many extended galaxies that are smaller than our cluster finding pixel scale of $12^{\prime\prime}$ (see \S 
 \ref{subsec:algorithm}) can still create spurious detections. To deal with this, we define a low-redshift sample of extended galaxies from the LS data with $z<0.5$, $m_z<17$ (AB), and with half-light radii ($R_{e}$) in the range $3.5^{\prime\prime}<R_{e}<15^{\prime\prime}$. We reject objects at $z>1$ for which the separation between \textit{WISE} and LS data ($s$) exceeds $2^{\prime \prime}$ if they lie within $2.5R_{e}$ of an object in the sample of low-redshift, extended galaxies.

\begin{figure}[htbp]
\centering
\includegraphics[width=0.9\columnwidth]{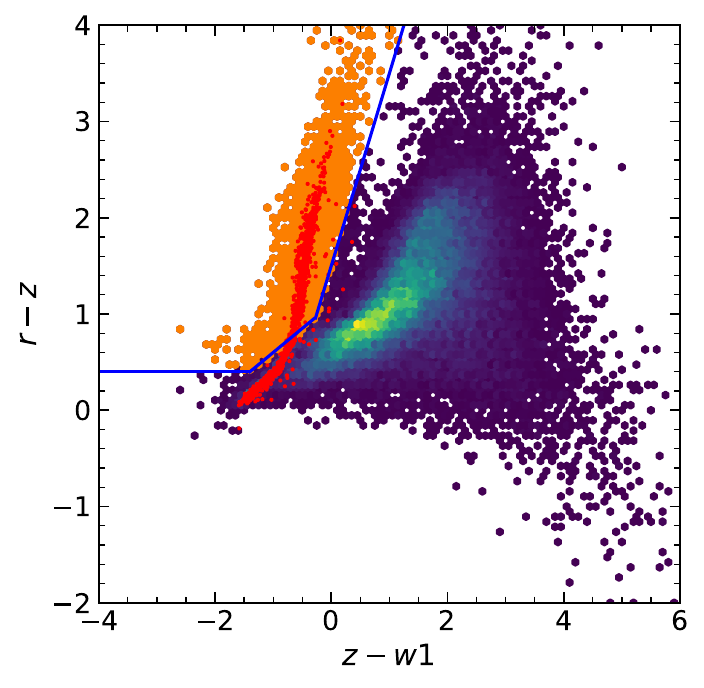}
\caption{Color-color plot showing the stellar locus in orange and red. The red points are objects for which the parallax is detected at $\geq5\sigma$ from $Gaia$ DR2 \citep{2018Gaia}. The color for the remaining objects indicates the density of objects on the plot. The blue line illustrates the equations used to select the orange locus \ref{eq:star}.}
\label{fig:starcut}
\end{figure}

In addition to the process above, we remove stars from the catalog as done in \cite{2021Decker}. Figure \ref{fig:starcut} demonstrates this cut. The stellar locus (orange) as well as the objects with $\geq 5\sigma$ parallax measurements in $Gaia$ DR2 \citep[red;][]{2018Gaia} are excluded from the input. The orange locus is selected by

\begin{equation}
\begin{split}
   \textit{r}-\textit{z} &> 0.5(\textit{z}-W1) + 1.1\;,\rm and\\
   \textit{r}-\textit{z} &> 2(\textit{z}-W1) + 1.5\;,\rm and\\
   \textit{r}-\textit{z} &> 0.4.
\end{split}
\label{eq:star}
\end{equation}

\section{Photometric Redshifts} \label{sec:photo-z}
Galaxy photometric redshifts were derived using the five-band optical/IR matched catalogs described above. Following a methodology similar to \citet{2006Brodwin}, we computed the $\chi^2$ surface in photometric redshift and galaxy template, using a subset of the empirical templates of \citet{2007Polletta} -- specifically, the Ell5, Ell13, S0, Sa, Sb, Sc, Sd, Spi4, and M82 templates --  plus the elliptical template of \citet{1980CWW} extended to the IR and UV using a $\tau = 1$ Gyr stellar population synthesis (SSP) model from \citet{2003Bruzual}.

The photometric redshift probability distribution functions (PDFs) are created by projecting $\chi^2$ along the redshift dimension.  No priors are used in this work. Detailed tests of the photometric zero points and effective wavelengths in all filters indicated that the calibration and characterization of both the LS and CatWISE2020 surveys are excellent, with no post hoc tweaks needed to optimize redshift estimation. More information on the photometric redshifts used in MaDCoWS2 are presented in Brodwin et al. (in prep).

\begin{figure}[htbp]
\centering
\includegraphics[width=0.9\columnwidth]{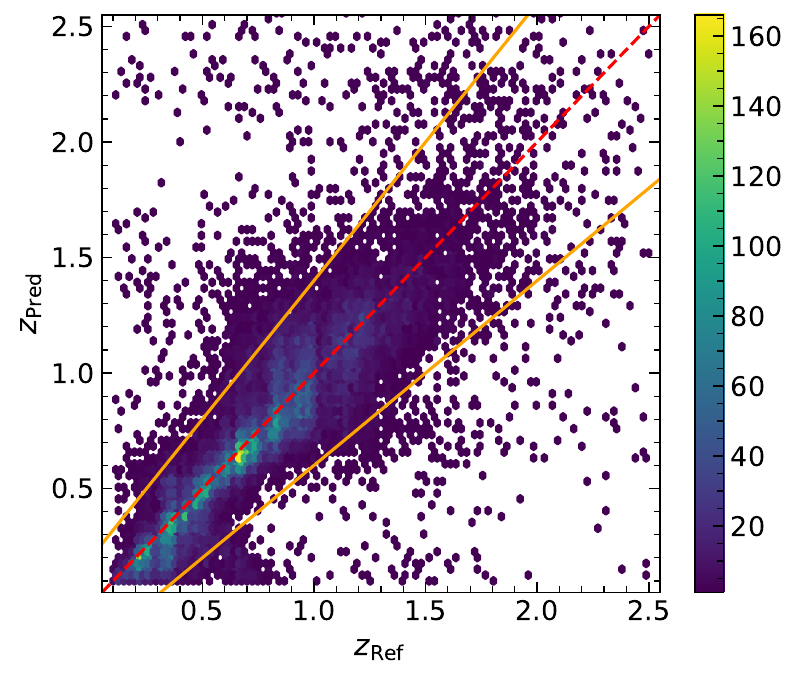}
\caption{Comparison between the photometric redshift derived in this work and redshift from the reference catalog. Color indicates the density of objects. The red line indicates a 1-to-1 line. The orange lines indicate $|\delta z/(1+z) |=0.2$. The outlier fraction is 8\% and $\sigma_z/(1+z)=0.075$.}
\label{fig:photz_qual}
\end{figure}

We compare our photometric redshifts to the redshifts of $31,981$ galaxies in the literature, which include redshifts from multiple spectroscopic catalogs and high-quality photometric redshifts from COSMOS. The spectroscopic sample of $11,705$ galaxies includes data from the IRAC Shallow Cluster Survey (ISCS, \citealt{2006Brodwin}), AGES \citep{2012Kochanek}, VVDS \citep{2013Le}, zCOSMOS \citep{2010Pozzetti}, and C3R2 \citep{2017Masters,2019Masters,2020Masters,2021Stanford}. Since the spectroscopic redshift sample is small at $z$ $>1.5$, we augment these spectroscopic redshifts with the COSMOS photometric redshift catalog \citep{2016Laigle,2021Razim}, which has uncertainties $\delta z /(1+z_{\rm Spec})<0.02$ where $\delta z\equiv (z_{\rm COSMOS}-z_{\rm Spec})$. Figure \ref{fig:photz_qual} illustrates the quality of the photometric redshifts at $z \geq 0.1$. The reference redshifts are matched by position within 2$^{\prime\prime}$. If we define the outliers to be objects with $\delta z/(1+z) >0.2$ and exclude them, the median of $\delta z/(1+z)$ is $\Tilde{\delta z}=-0.017$. The standard deviation is $\sigma_z/(1+z) = 0.075$. The outlier fraction $\eta$ is $8\%$. 

In \S \ref{sec:data prep}, we considered the distribution of matching separation ($s$). The probability that a match is real at a radius affects the PDF of a source redshift, as it tells us the chance that a source will contain all five bands of photometry. Using a specific matching radius could exclude a real match outside of that radius. Weighing this effect, the PDF then can be calculated by 

\begin{equation}
\begin{multlined}
    \text{PDF} = \text{erfc}(s,\mu,\sigma)\times \text{PDF}_{\text{match}}
\end{multlined}
\label{eq:frac}
\end{equation}
where erfc is the complementary error function describing the probability that a match is real and $s$ is the separation between a match. This function is used when the probability of a real match is $<90\%$ (see y-axis in the bottom panel of Figure \ref{fig:sep}). Galaxies with the probability of a real match $< 50\%$ are removed from the cluster search. 

\section{Cluster Finding} \label{sec:cluster finding} 
The pipeline for MaDCoWS2 consists of the main cluster finding algorithm, called PZWav, and a post-processing section. We use the PZWav algorithm for cluster finding and develop a tailored pipeline optimized for MaDCoWS2 to search over wide areas with an appropriate tiling strategy. The post-processing section takes the initial detections from PZWav and generates the final columns for the catalog. Details of the pipeline implementation are given below.

\subsection{PZWav} \label{subsec:algorithm}
PZWav \citep{2014Gonzalez,2019Euclid} searches for galaxy clusters using their overdensity. This algorithm is based on the method used initially in the \textit{Spitzer} IRAC Shallow Cluster Survey (ISCS) in \cite{2008Eisenhardt} and the \textit{Spitzer} IRAC Deep Cluster Survey (IDCS) in \cite{2012Stanford}. It has also been selected by the \textit{Euclid} mission as one of the two cluster detection algorithms \citep{2019Euclid} and has been used by \citet{2023Werner} for the S-PLUS survey. 

For MaDCoWS2, the code has been optimized for our data. PZWav takes the photometric redshifts and sky positions of galaxies as input. It generates a 3D data cube in RA, DEC, and redshift with pixels of width $12^{\prime\prime}$ in RA, and DEC, and $0.06$ in redshift. A 3D density map is then created by adding galaxies to the data cube at the appropriate (RA, DEC), with the PDF used to weight their contribution as a function of redshift. The weights of galaxies in each density map are calculated from the integral of the galaxies PDF within the redshift slice. The density maps are normalized to have the same mean value as the density map at $z$ $=1$. We smooth these density maps by applying a difference-of-Gaussian smoothing kernel with a fixed physical size matched to the typical size of cluster cores. The inner and outer kernel standard deviation sizes are $400$ kpc and $2$ Mpc. 

For all the maps, we apply masks (see \S \ref{subsec:contamination}) to minimize spurious detections. These include masks for bright stars and large galaxies. We detect clusters as overdensities with signal-to-noise ratio S/N $\geq5$. These overdensities are peaks in the data cube detected in three dimensions. The position of the peak defines the cluster centroid. If multiple peaks exist within 1 Mpc and $\pm$0.12 photometric redshift of one another, only the strongest peak is selected. The algorithm calculates a uniform noise threshold from a set of random density maps in bootstrap simulations. The random maps are created by randomly shuffling the PDFs relative to their positional information. The S/N is determined from the amplitude of the peak in the smoothed density maps and the fluctuations in the random maps. The PZWav code is capable of computing S/N based upon the assumption of either Gaussian or Poisson background noise. For this paper, S/N hereafter refers to calculations based upon the assumption of a Poisson noise distribution, which generally leads to lower S/N values than the Gaussian approximation. In the catalog we also include the corresponding Gaussian S/N (denoted S/N$_G$) for completeness and to enable comparison with any futures catalogs based upon PZWav that opt to use the Gaussian S/N. The redshifts of the cluster candidates are refined by computing the $\sigma-$clipped median photometric redshift of all galaxies that lie within $750$ kpc of the peak centroids for which $z_{\rm phot}$ lies within $\pm$ three redshift slices ($0.18$) from the slice containing the peaks. The redshift errors are calculated from $\sigma_{z_{\rm Spec}}\times(1+z_{\rm phot})$ with $\sigma_{z_{\rm Spec}}=0.029$. We derive $\sigma_{z_{\rm Spec}}$ by comparing the photometric redshifts of MaDCoWS2 clusters to spectroscopic redshifts in existing catalogs (see \S \ref{subsec:Redshift comparison}).

We occasionally find multiple detections along the same line of sight. This may be a chance projection, but may also indicate that the background candidate is spurious. The background candidate may arise from secondary peaks in the PDFs of galaxies in the lower redshift cluster. For each higher redshift candidate which lies behind a lower-redshift cluster, we identify foreground clusters. A detection is flagged as a foreground if it lies within a projected $500$ kpc of the more distant candidate and has a redshift at least $0.12$ (two bins) lower than that of the candidate. We then add a column, Name$_{\rm fg}$, to Table \ref{tab:sample2} which contain the name of the foreground clusters. This column is generated in the post-processing section of the pipeline. 

We further filter the cluster catalog to remove candidates whose centroids lie on the edges of masked regions of our smoothed density map. Specifically, we require our cluster candidate to have the fraction of area masked $f_{\rm area}<0.4$ within $30^{\prime\prime}$ and $120^{\prime\prime}$ of the cluster position. This process is also performed in the post-processing section of our pipeline.

\subsection{Tiling} \label{subsec:tiling}
\begin{figure*}[htbp]
\centering
\subfigure{\includegraphics[width=0.9\textwidth]{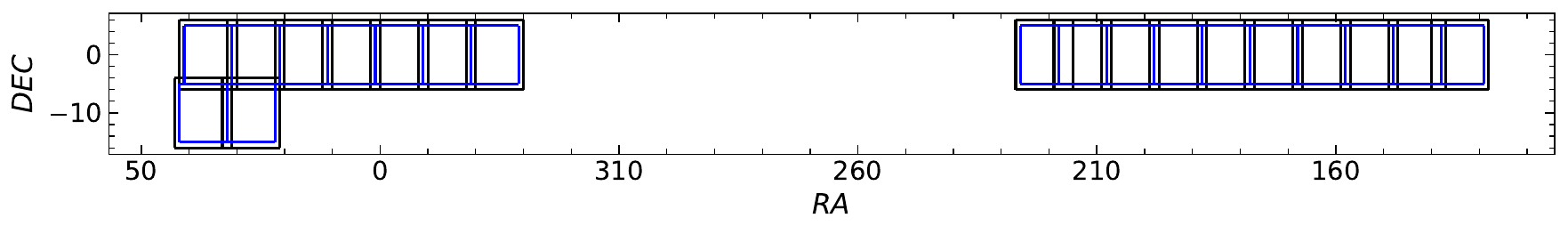}}
\caption{The tiling for this work. We run PZWav on each of the black rectangles independently, and merge the results using the tiling from the blue rectangle.} 
\label{fig:tiles}
\end{figure*}

To enable our wide area survey, we divided our survey region into multiple $12^\circ\times12^\circ$ tiles which overlap one another by at least 1$^\circ$ on each edge \footnote{These tiles are different from the catalog matching tiles described in \S \ref{subsec:Catalog Matching}}. These tiles are shown as black rectangles in Figure \ref{fig:tiles}. We use the bisector of the overlap region to demarcate the division between tiles when selecting candidates. Thus, the effective area of the search in each tile is $10^\circ\times10^\circ$. We run PZWav on each tile independently. The catalogs for each tile are then filtered to include only cluster detections that lie within the region defined by the midpoint of the overlap region with adjacent tiles (only objects within each $10^\circ\times10^\circ$ tile are included), and these filtered catalogs are merged. For the tiles on the two edges in RA, the overlap is adjusted so that these tiles are the same size as those in the middle, and the overlap region is bisected when creating the merged catalogs. For all tiles at the survey edges, we filter out the clusters within $1^\circ$ of the edges. These filtered tiles are shown as blue rectangles in Figure \ref{fig:tiles}. We merged detections from each blue tile to create the final catalog.

Duplicated detections can remain due to structures that are detected around the midpoint of the overlap region. We handle this by applying the same algorithm as the one used before the redshift refinement in \S \ref{subsec:algorithm}. The algorithm selects the higher S/N detection when there are pairs of detection within $1$ Mpc and $0.12$ in photometric redshift.

\subsection{Masking and Candidate Rejection} \label{subsec:contamination}
MaDCoWS2 relies upon galaxy overdensities to find clusters, and therefore false cluster detections can arise from overdensities of spurious \textit{WISE} sources. Spurious \textit{WISE} sources will typically lack optical counterparts (or have faint optical counterparts), and as optical dropouts, the photo-z PDF will peak at high redshift. If these artifacts are spatially clustered, they will yield false detections of high-redshift clusters -- i.e., the maximal contamination will be in the most interesting redshift interval. It is thus critical to minimize the impact of this contamination by not only filtering the catalogs (see \S \ref{sec:filtering}) but also masking regions where such contamination is expected.

\subsubsection{Fragmented Galaxies}\label{sec:Fragmented Galaxies}
\begin{figure*}[htbp!]
    \centering
        \includegraphics[width=0.9\columnwidth]{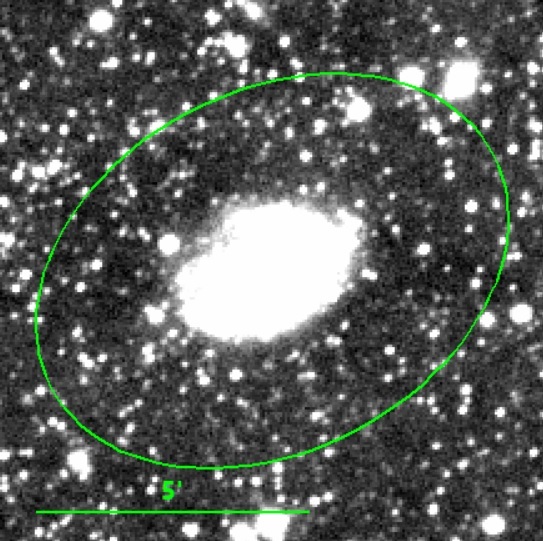} 
        \includegraphics[width=0.9\columnwidth]{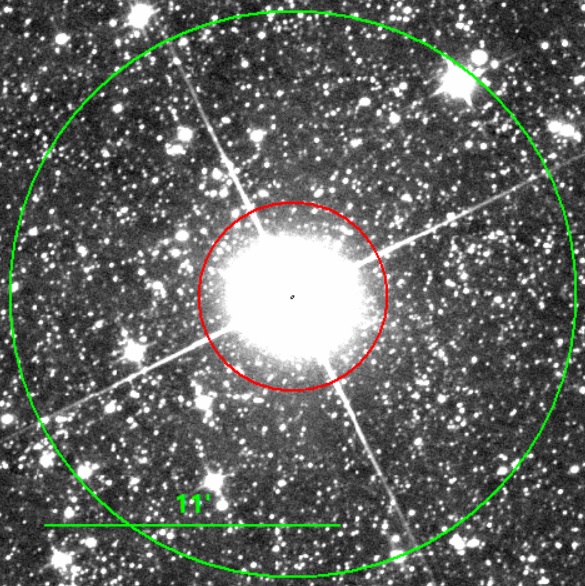}  
        \caption{Left: NGC 895 in W1 unWISE coadd image $(10'\times10')$. The SGA catalog galaxy mask is shown as a green ellipse. Right: TIC 35723942 \citep{2019Stassun} with AllWISE $W1=3.54$ in an W1 unWISE coadd image $(22'\times22')$. The MaDCoWS2 star mask is shown as a green circle. The red circle shows the size of the mask in the original MaDCoWS program} 
        \label{fig:contamination}
\end{figure*}

One particular concern is that the CatWISE2020 photometry is derived using \texttt{crowdsource}, which treats all objects as point sources. Sources that are spatially resolved with \textit{WISE} are modeled as a superposition of point sources. This fragmentation thus results in a single central source that is matched with the optical catalog, surrounded by unmatched fragments that are assigned to high redshift. We present an example of galaxy with this type of artifact in Figure \ref{fig:contamination}. We mitigate this issue via conservative masking of known large, extended galaxies. We create two different sets of elliptical masks.

The first set of galaxy masks utilizes the Siena Galaxy Atlas 2020 \citep[SGA;][]{2021Moustakas}, which contains over 300,000 large, nearby galaxies. The models from the SGA catalog come from multi-band surface brightness profiles derived from fitting elliptical isophotes. Each model comes with the semi-major axis radius at the $26$ mag arcsec$^{-2}$ $r$-band isophote $(D26)$, position angle $(PA)$, and minor-to-major axis ratio $(BA)$. In this set of galaxy masks, we mask galaxies for which $(D26)$ is larger than $30^{\prime\prime}$. The mask is defined as an ellipse with a semi-major axis $D26$, a position angle $PA$, and a minor-to-major axis ratio $BA$. An example of this mask is shown as the green ellipse in Figure \ref{fig:contamination} (left). We apply these masks to the density cube in the PZWav section of our pipeline.

The second set of galaxy masks uses the galaxy models from the LS catalog. The Tractor code in LS finds the best fit for galaxies from five different models including round exponential galaxies with a variable radius, deVaucouleurs profiles (elliptical galaxies), exponential profiles (spiral galaxies), and Sersic profiles. Each LS model comes with half-light radius of galaxy model $R_{e}$ and ellipticity components ($\epsilon_1\;\text{and}\;\epsilon_2$). We require galaxies used for this set of masks to have $R_{e}>4"$, maximum LS $z$-band magnitude of $19$, and maximum LS photometric redshift \citep{2021Zhou} of $0.5$. The sizes of these masks are $2 R_{e}$ of the selected galaxies. In the post-processing section of our pipeline, we reject galaxy cluster candidates at $z$ $>1$ that are in these masked regions.

We use the combination of the LS and SGA masks because the SGA catalog is more effective for the largest nearby galaxies, for which the LS models may not be robust, while the LS extends the masking to smaller size and fainter galaxies. The SGA masking is applied to the density cube during the search to avoid contamination at all redshifts. The LS masking is designed to avoid spurious detections at $z$ $>1$ due to fragmentation of \textit{WISE} sources, and is applied in the post-processing phase. We apply LS masking only at $z$ $>1$ because many LS masks are made from galaxies small enough to be members of $z$ $<1$ galaxy clusters.

\subsubsection{Bright Stars} 
Section \ref{sec:filtering} discussed the removal of artifacts from bright stars using flags from the \textit{WISE} catalog. Because not all artifacts may be removed by catalog flags, we also generate masks around bright stars in a fashion similar to what was employed for the original MaDCoWS program.
For each bright star, a circular mask is created using the W1 magnitude in AllWISE and the size$-$magnitude relation for scattered-Light halos in \S IV.4.g.ii.1.a of \cite{2012Cutri}. The star mask used in the original MaDCoWS program used scaled magnitudes derived from W1scaled $\equiv$ W1$-\log_{10}(\text{W1cov})$ where W1cov is the mean coverage depth in W1 (typically 12). We used the same equation but extended the masks based on visual inspection of unWISE W1 images. 

The goal of our star masks is to avoid possible artifacts close to bright stars. Thus, we design masks to cover the halo and diffraction spikes of bright stars. The minimum and maximum mask radii are $4.5$ and $459.2$ pixels ($54^{\prime\prime}$ and $5511^{\prime\prime}$). From visual inspection of unWISE W1 images, we found that larger masks than the ones given by the original MaDCoWS program provide superior performance for the search. We use masks that are larger by factors of $6$, $3.5$ and $1.5$ for stars with W1scaled$<2$, $2\leq$W1scaled$\leq4$, and W1scaled$>4$, respectively. An example of this mask is shown in Figure \ref{fig:contamination} (right) as a green circle with the original mask from the MaDCoWS program as a red circle.

We supplement W1scaled from AllWISE with W1 magnitudes (W1$_{\rm AS}$ $<6.5$) from the \textit{WISE} All-Sky data release \citep{2012Cutri_a} because it provides more robust photometry than AllWISE for these bright stars \citep{2013Cutri}. In the case that All-Sky objects duplicate AllWISE objects, we use the brighter magnitude from the two catalogs to create our bright star masks.

Occasionally, some stars possess diffraction spikes or halos that extend further than the default mask. In our search area, there are $8$ stars for which this is the case (see Table \ref{Tab:stars}). We use W1$_{\rm AS}$ to create masks based on the size$-$magnitude relation, and increased the mask radii from the relation by a factor of $5$. We also extend the mask radii by $25^{\prime}$ to ensure they cover the extended diffraction spikes and halos of these bright stars. We use rectangular masks for the diffraction spikes of these stars.

\begin{deluxetable}{ccccc}
\tablecaption{Bright Stars Requiring Additional Masking \label{Tab:stars}}
\tablehead{\colhead{Name} & \colhead{RA} & \colhead{Dec} & \colhead{W1$_{\rm AS}$} & \colhead{Radius} \\ 
\colhead{} & \colhead{(Deg)} & \colhead{(Deg)} & \colhead{} & \colhead{(Arcmin)}}
\startdata
30 Psc & 0.4910 & -6.0138 & 2.242 & 48\\
Omicron Cet & 34.8375 & -2.9766 & 1.870 & 51\\
RT Vir & 195.6583 & 5.1850 & 2.089 & 49\\
Delta Vir & 193.9002 & 3.3966 & 2.022 & 49\\
BK Vir & 187.5884 & 4.4166 & 2.138 & 49\\
SW Vir & 198.5176 & -2.8056 & 2.077 & 49 \\
Alpha Her & 258.6620 & 14.3927 & 1.563 & 53 \\
19 Psc & 356.5979 & 3.4867 & 2.134 & 49 \\
\enddata
\tablecomments{The halos or diffraction spikes of these stars extend beyond our default mask sizes and require additional manual masking.}
\end{deluxetable}

\section{The Catalog} \label{sec:Basic characteristics}
We present a catalog of $22,970$ cluster candidates from MaDCoWS2 with S/N $>5$. The redshift range of this cluster catalog is $0.1<z<2$. A total of $5114$ of the candidates lie at $z$ $>1$ and $1312$ at $z>1.5$. $3775$ candidates are rejected based on the fraction of area masked (see \S \ref{subsec:algorithm}) to obtain this catalog. $80$ candidates have foreground cluster detections. We set the lower bound of our search to $z=0.1$ because at lower redshifts the comoving volume is small and the photometric redshift uncertainties are large ($\sigma_z/(1+z)\sim z$). At high redshift, while we detect cluster candidates at $z>2$, we set $z=2$ as the upper bound for the current catalog, as we are still assessing the fidelity of the sample at higher redshift.

The catalog contains name, RA, DEC, photometric redshift (and its error), S/N from Gaussian noise (S/N${_G}$), S/N from Poisson noise (S/N), and the names in the literature of each cluster candidate (and numbers assigned to the referred works). We additionally provide spectroscopic redshifts derived from external catalogs along with their references. Column Name$_{fg}$ (see \S \ref{subsec:algorithm}) is also included for a foreground detection. 

\begin{deluxetable}{ccc}
\tablecaption{The sky coverage of MaDCoWS2 DR1.\label{tab:coveragetable}}
\tablehead{\colhead{Field} & \colhead{RA} & \colhead{DEC} } 
\startdata
$1$ & $331^\circ\;\text{to}\;41^\circ$ & $-5^\circ\;\text{to}\;5^\circ$ \\
$2$ & $22^\circ\;\text{to}\;42^\circ$ & $-15^\circ\;\text{to}\;-5^\circ$ \\
$3$ & $129^\circ\;\text{to}\;226^\circ$ & $-5^\circ\;\text{to}\;5^\circ$ \\
\enddata
\tablecomments{Fields 1 and 3 cover the Fall equatorial field of the HSC DR3 \citep{2022Aihara} while Field 2 covers the Spring equatorial field.}
\end{deluxetable}

\begin{figure*}[htbp]
\centering
\subfigure{\includegraphics[width=0.9\textwidth]{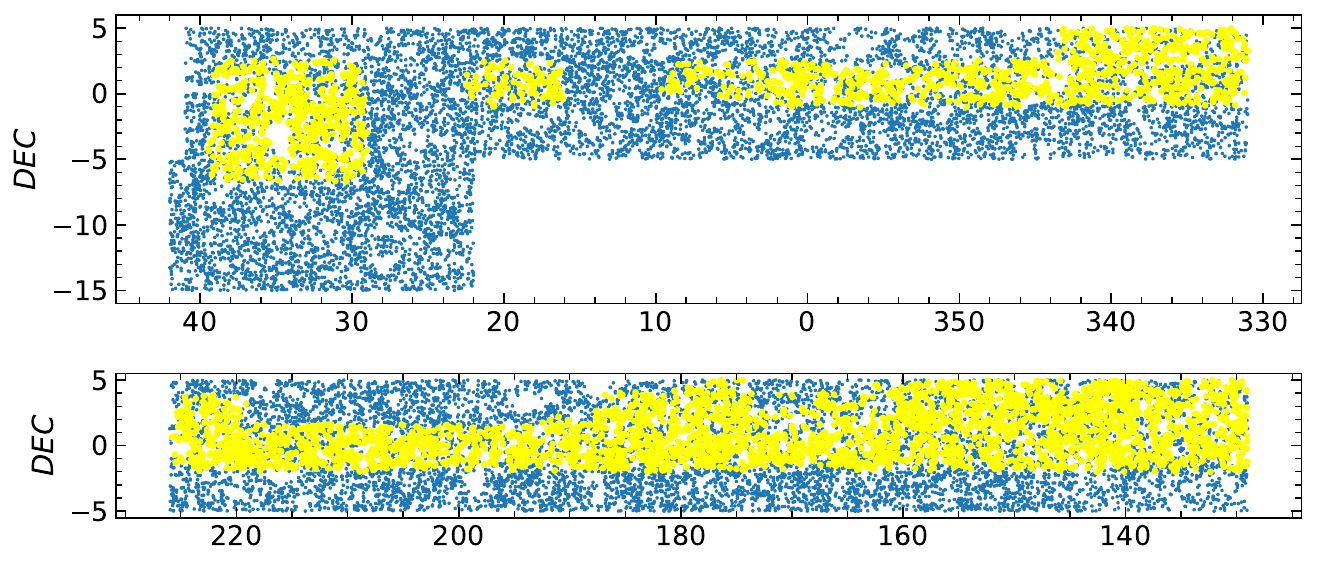}}
\subfigure{\includegraphics[width=0.9\textwidth]{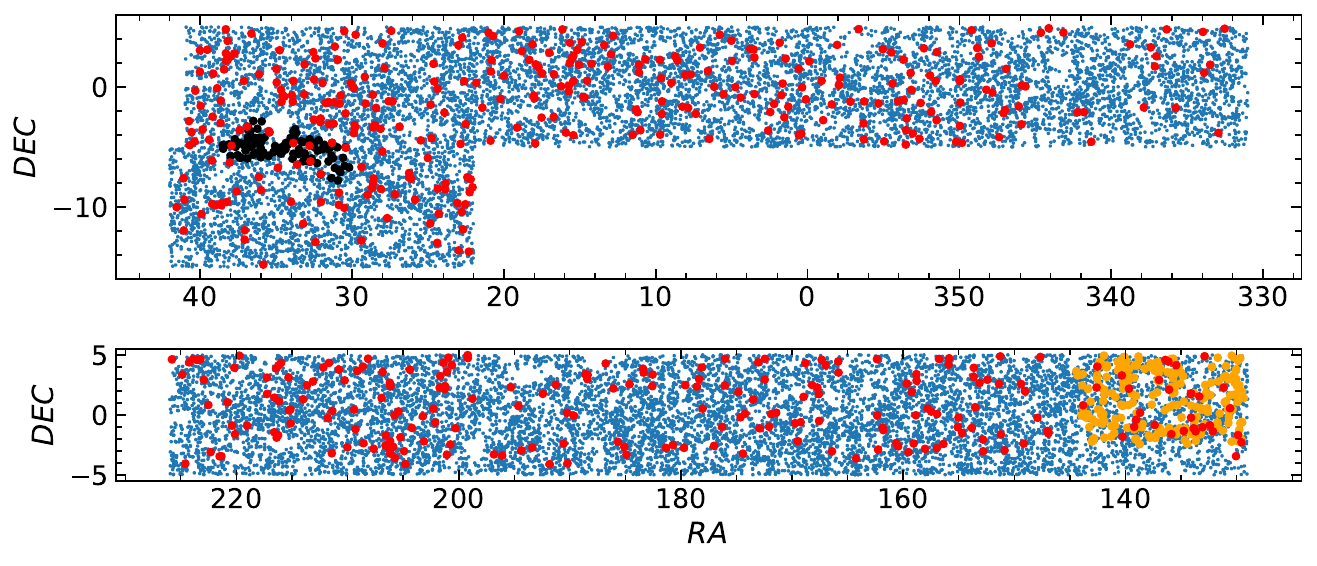}}
\caption{Sky distribution of cluster candidates (blue) in the HSC wide Fall equatorial field (the first panel from the top) and the HSC wide Spring equatorial field (the second panel). Yellow points indicate HSC CAMIRA clusters \citep{2018Oguri}. The third and fourth panels show the spatial distribution of ACT clusters \citep{2021Hilton}, XXL clusters \citep{2018Adami}, and eFEDS clusters \citep{2022Liu} using red, black, and orange points, respectively.} 
\label{fig:spatdist}
\end{figure*}

We show the spatial distribution of this survey in Figure \ref{fig:spatdist} and Table \ref{tab:coveragetable}. The area of the search is approximated to be $1461$ deg$^2$ after removing the masked area ($1838$ deg$^2$ without masking). This area covers the equatorial fields of HSC wide DR3 \citep{2022Aihara}. The CAMIRA cluster catalog from HSC DR3 is shown as yellow points in Figure \ref{fig:spatdist} (the catalog is updated from DR2 by the HSC collaboration using the same method as in \cite{2018Oguri}). The ACT clusters \citep{2021Hilton}, the XXL clusters \citep{2018Adami}, and the eROSITA Final Equatorial-Depth Survey (eFEDS) clusters \citep{2022Liu} are the red, black, and orange points, respectively. These four catalogs are among the external catalogs we compare with in \S \ref{sec:survey characterization}. Their descriptions are presented in \S \ref{subsec:comparison}.

\begin{figure}[htbp]
\centering
\includegraphics[width=0.9\columnwidth]{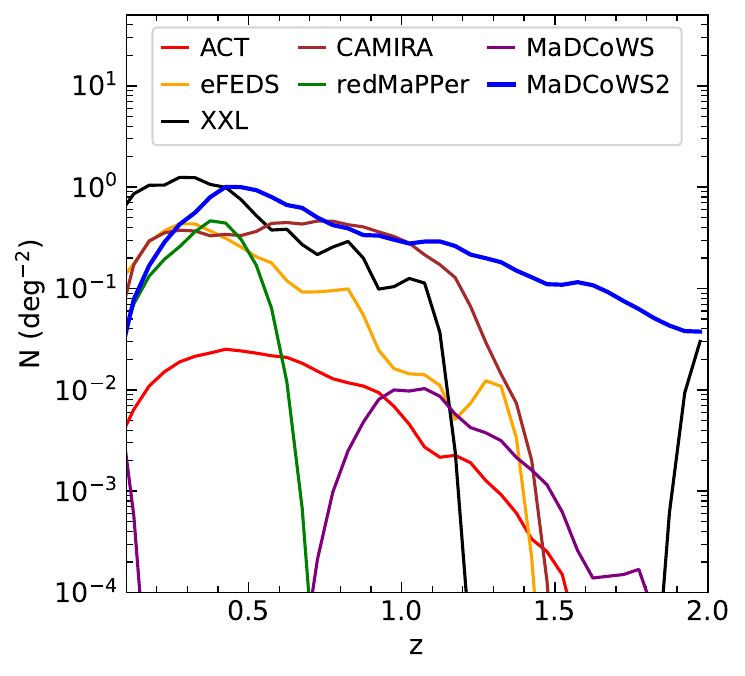}
\caption{Photometric redshift distribution of MaDCoWS2 and external catalogs normalized by the coverage areas. The distributions are smoothed by a Gaussian filter with $0.15$ redshift radius and standard deviation of $0.75$.} 
\label{fig:zdist2}
\end{figure}

Figure \ref{fig:zdist2} show the estimated smoothed redshift distributions of MaDCoWS2 and external catalogs normalized by the coverage areas of the catalogs. The MaDCoWS2 redshift distribution peaks at $z\sim0.5$, declining at lower redshift due to the smaller comoving volume and decreasing at higher redshift due to a combination of cluster mass evolution and lower S/N at fixed cluster mass. MaDCoWS2 provides a comparable number of detections per deg$^2$ to ACT, eFEDS, and CAMIRA at the lowest redshifts, but a greater projected density of candidates at higher redshift. 

\begin{figure}[htbp]
\centering
\includegraphics[width=0.9\columnwidth]{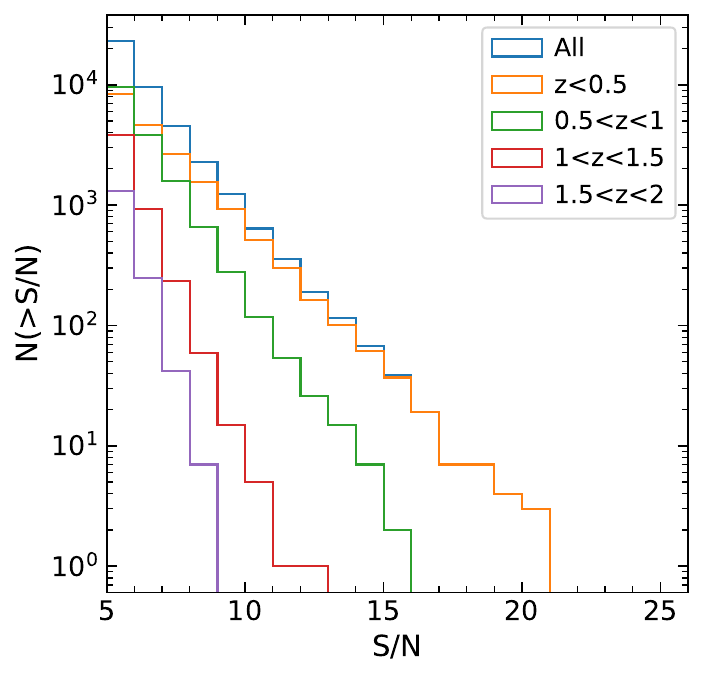}
\caption{Cumulative S/N distribution for MaDCoWS2 cluster candidates as a function of redshift.}
\label{fig:SNRdist}
\end{figure}

We show the distribution of the S/N of the candidates in Figure \ref{fig:SNRdist} in a cumulative histogram. While the highest S/N clusters are found at $z$ $<0.5$, there exist candidates with $S/N>8$ even at $z>1.5$.

Figure \ref{fig:pic1} show images of the 3 highest S/N cluster candidates in different redshift ranges, $z<0.5,\;0.5<z<1,$ $1<z<1.5,$ $1.5<z<1.75$, and $1.75<z<2.0$. We use the images created from the $g$, $r$, and $z$ bands of LS DR9 for these candidates. For candidates at $z>1$, we also display W1 images from unWISE \citep{2014Lang} to highlight the galaxies that become faint at optical wavelengths but appear at infrared wavelengths. Galaxies at $z>1.5$ are faint at optical wavelengths due to the large k-corrections as the Balmer break passes out of the optical, and thus typically are too faint to see in the LS DR9 images. These high redshift galaxies are brighter in infrared, so we use W1 images to display these galaxies. Each panel of Figure \ref{fig:pic1} contains the name of the cluster candidate, its S/N, its photometric redshift, and its name and spectroscopic redshift from the external catalog (if available). The red circles on the \textit{WISE} images are galaxies with integrated PDF $>0.2$ where the integrated PDF is defined by the integration of the PDF from $-0.06(1+z_{\rm cluster})$ to $+0.06(1+z_{\rm cluster})$. 

\begin{figure*}[htbp]
\begin{center}
\begin{tabularx}{\textwidth}{ccc}
\includegraphics[width=0.3\textwidth]{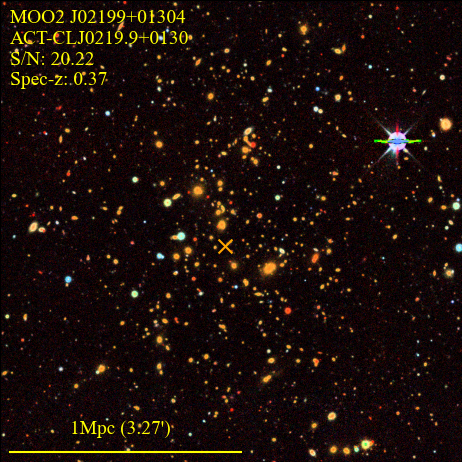} & 
\includegraphics[width=0.3\textwidth]{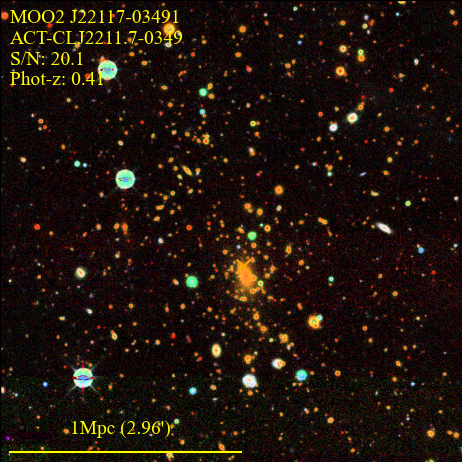} &
\includegraphics[width=0.3\textwidth]{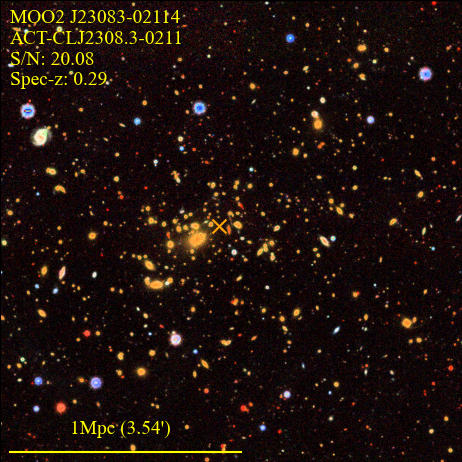}
\end{tabularx}
\begin{tabularx}{\textwidth}{ccc}
\includegraphics[width=0.3\textwidth]{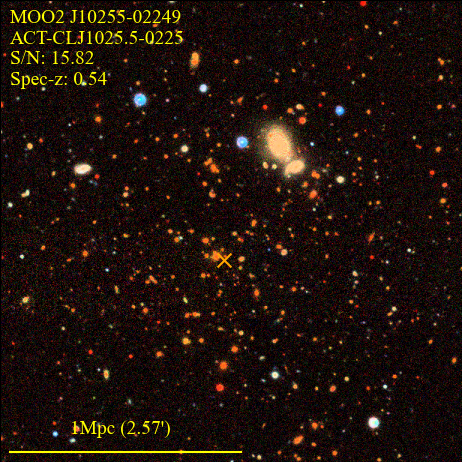} & 
\includegraphics[width=0.3\textwidth]{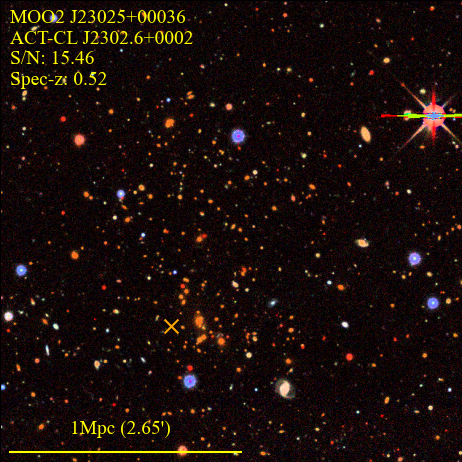} &
\includegraphics[width=0.3\textwidth]{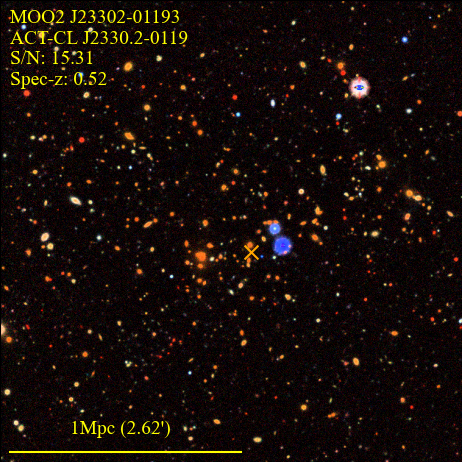}
\end{tabularx}
\begin{tabularx}{\textwidth}{ccc}
\includegraphics[width=0.3\textwidth]{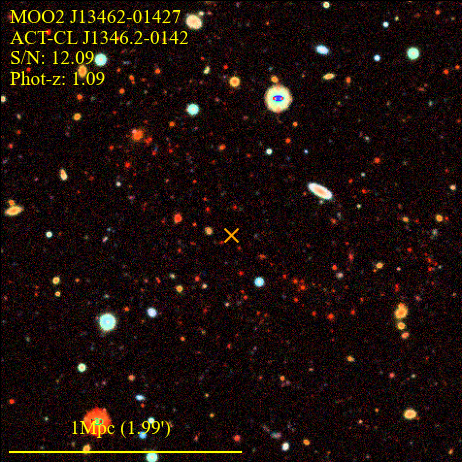} & 
\includegraphics[width=0.3\textwidth]{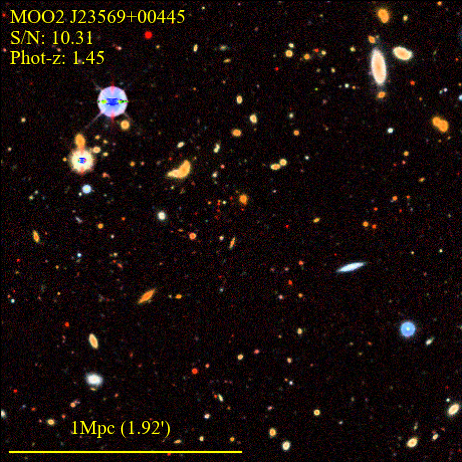} &
\includegraphics[width=0.3\textwidth]{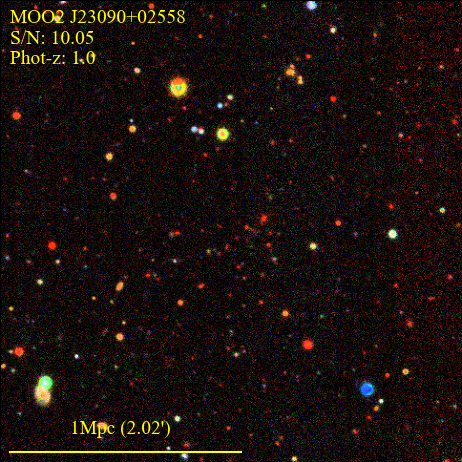}
\end{tabularx}
\begin{tabularx}{\textwidth}{ccc}
\includegraphics[width=0.3\textwidth]{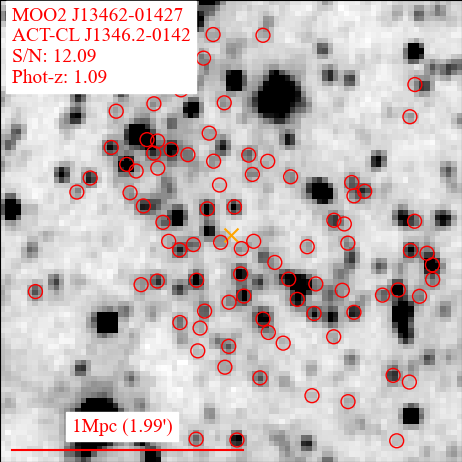} & 
\includegraphics[width=0.3\textwidth]{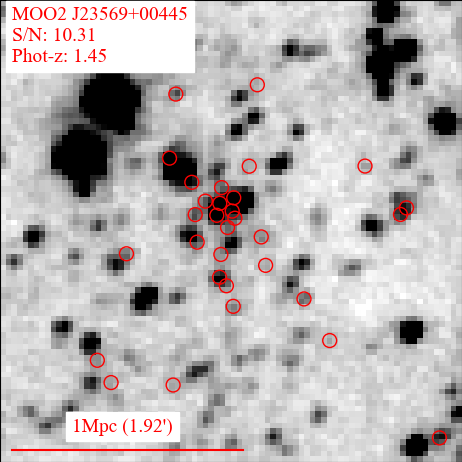} &
\includegraphics[width=0.3\textwidth]{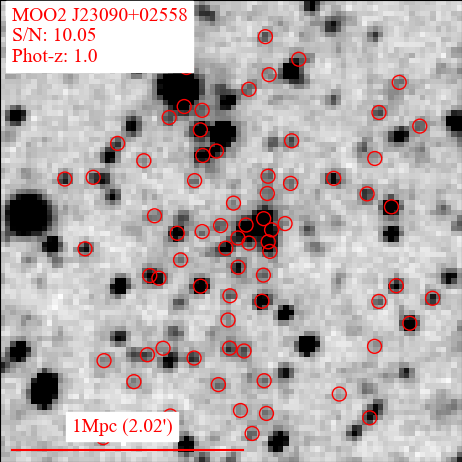}
\end{tabularx}
\caption{\label{fig:pic1} LS DR9 \textit{grz} images of three clusters candidates at $z$ $<0.5$ (first row), $0.5<$ $z$ $<1$ (second row), and $1<$ $z$ $<1.5$ (third row). The size of the images is $2 \text{~Mpc} \times 2 \text{~Mpc}$. Each row displays the highest S/N candidates in their respective redshift range. The fourth row displays the same candidates as those in the third row but in the unWISE W1 band. Orange Xs denote the centroid of ACT clusters. The red circles are objects with integrated PDF $>0.2$. In each panel, we provide the name of the cluster candidate, the name of its ACT counterpart, our S/N, and our photometric redshift (or spectroscopic redshift if available from the external catalog.)}
\end{center}
\end{figure*}

\begin{figure*}
\addtocounter{figure}{-1}
\begin{center}
\begin{tabularx}{\textwidth}{ccc}
\includegraphics[width=0.3\textwidth]{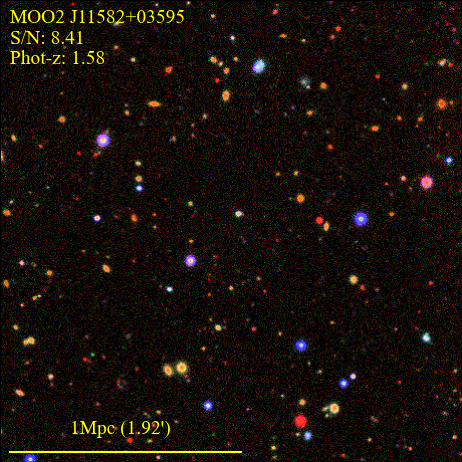} & 
\includegraphics[width=0.3\textwidth]{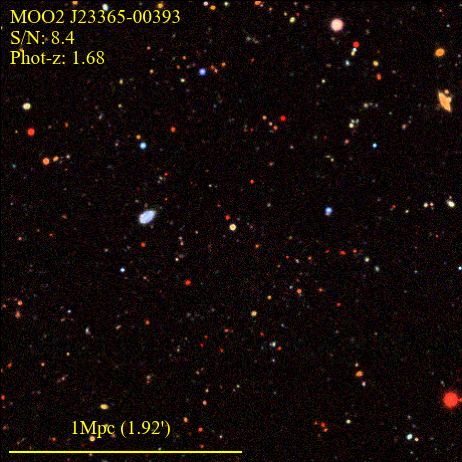} &
\includegraphics[width=0.3\textwidth]{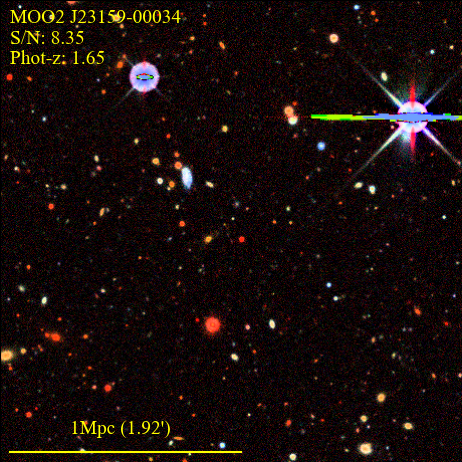}
\end{tabularx}
\begin{tabularx}{\textwidth}{ccc}
\includegraphics[width=0.3\textwidth]{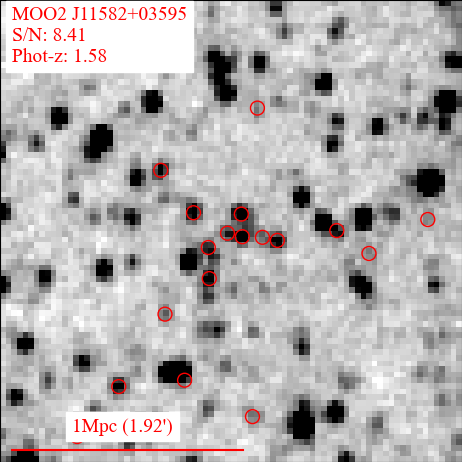} & 
\includegraphics[width=0.3\textwidth]{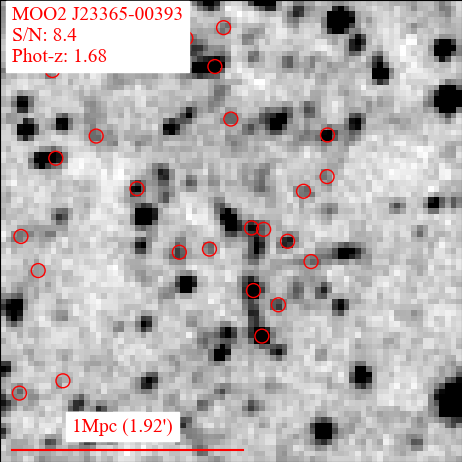} &
\includegraphics[width=0.3\textwidth]{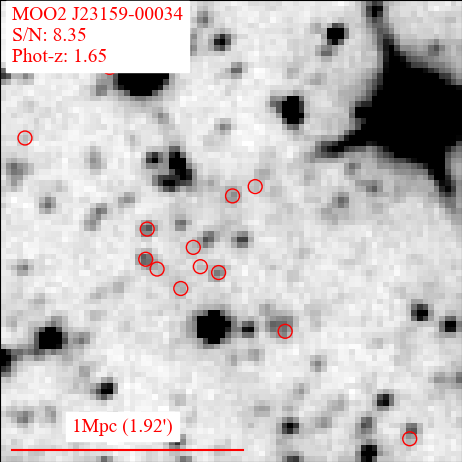}
\end{tabularx}
\begin{tabularx}{\textwidth}{ccc}
\includegraphics[width=0.3\textwidth]{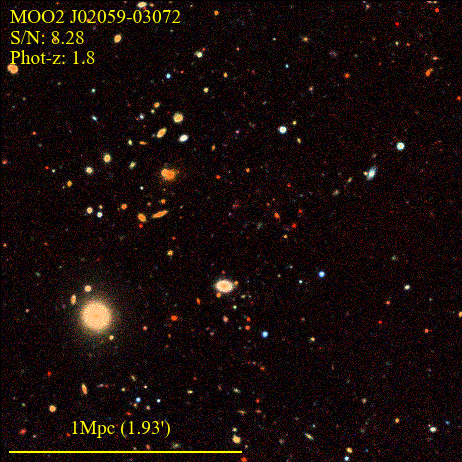} & 
\includegraphics[width=0.3\textwidth]{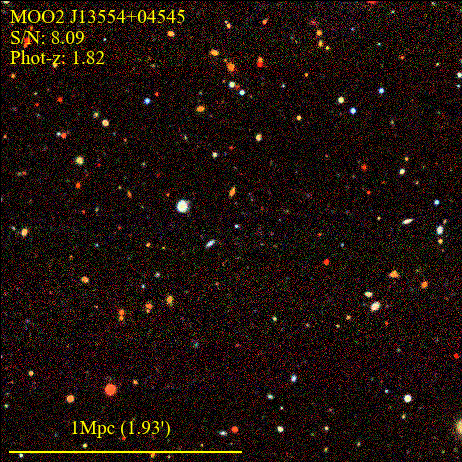} &
\includegraphics[width=0.3\textwidth]{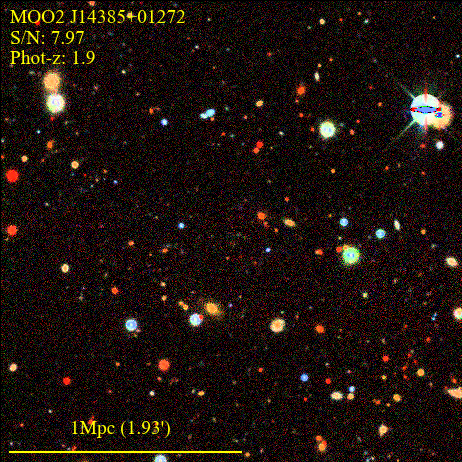}
\end{tabularx}
\begin{tabularx}{\textwidth}{ccc}
\includegraphics[width=0.3\textwidth]{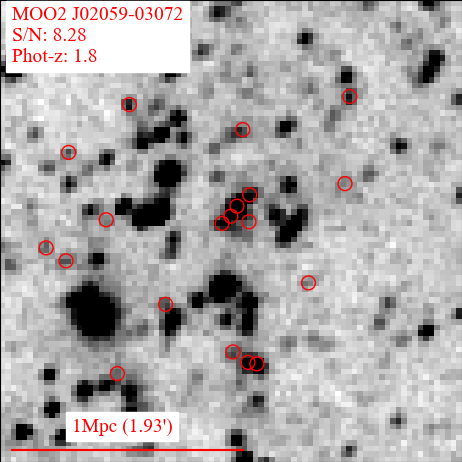} & 
\includegraphics[width=0.3\textwidth]{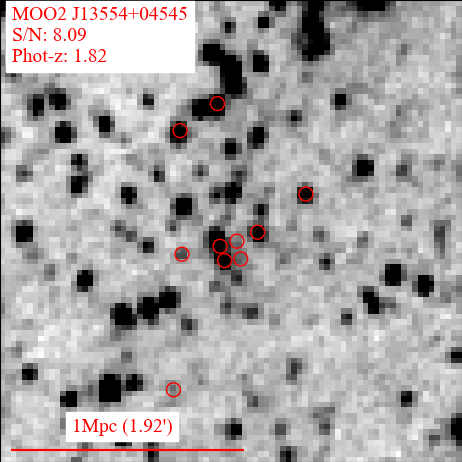} &
\includegraphics[width=0.3\textwidth]{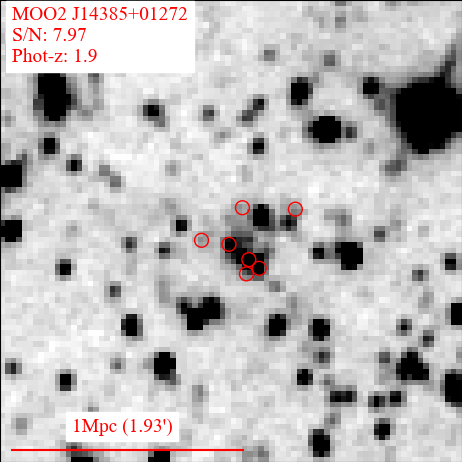}
\end{tabularx}
\caption{Continued) LS DR9  \textit{grz} and unWISE W1 images of top three highest-S/N clusters candidates at $1.5<$ $z$ $<1.75$ (first and second row) and at $1.75<$ $z$ $<2.0$ (third and fourth row).} 
\end{center}
\end{figure*}

\section{Survey Characterization} \label{sec:survey characterization}
We match the cluster candidates from our catalog to other clusters catalogs in our search area to assess the fidelity of our search. 

The ACT Sunyaev–Zel’dovich effect cluster survey \citep{2021Hilton} covers this region, as do the
XXL \citep{2018Adami} and the eFEDS \citep{2022Liu} X-ray catalogs.
There also exist multiple optical/infrared-based cluster searches that overlap with this region. These include
the redMaPPer catalog \citep{2016Rykoff}, the HSC CAMIRA cluster catalog \citep{2018Oguri}, and the MaDCoWS catalog \citep{2019Gonzalez}.  Each of these catalogs is described in greater detail below. 

\subsection{Existing Cluster Catalogs} \label{subsec:comparison}
\subsubsection{ACT} \label{subsec:ACT}
 ACT DR5 detects galaxy clusters using observations of the SZ effect at $98$ GHz and $150$ GHz over $13,211$ deg$^2$ of the sky. The catalog consists of $4195$ optically confirmed galaxy clusters with S/N$>4$ at $0.04<z<1.91$. The $90\%$ completeness of the catalog for S/N$>5$ is at $M_{500c}>3.8\times10^{14}M_{\odot}$, assuming a relation between the SZ signal and the mass calibrated from X-ray observations \citep{2010Arnaud}. ACT DR5 uses optical/IR data from external catalogs (including \textit{WISE}) to confirm their cluster detections. The spectroscopic redshifts also are derived from these catalogs. We refer to \cite{2021Hilton} for more information about the ACT DR5 catalog.
 
\subsubsection{eFEDS} \label{subsec:efed}
eFEDS is a $0.2-2.3$ keV X-ray survey covering $140$ deg$^2$ with an average exposure time of $1.3$ ks by eROSITA \citep{2012Merloni,2021Predehl}, which is onboard the \textit{Russian-German Spectrum-Roentgen-Gamma (SRG)} satellite \citep{2021Sunyaev}. eFEDS clusters use the Multi-Component Match filter (MCMF) algorithm \citep{2018Klein,2019Klein} to optically confirm their detections and derive the redshift of the clusters. eFEDS uses optical imaging from the DESI LS \citep{2019Dey} and the HSC Subaru Strategy Program \citep{2018Aihara}. \cite{2022Klein} provides a full description of the analysis. Each source in eFEDS comes with a detection likelihood $L_{\rm det}$ and an extent likelihood $L_{\rm ext}$. We select objects with $L_{\rm det} > 5$ and $L_{\rm ext} > 6$ which results in $80 \%$ real clusters based on analysis from simulations in \cite{2020Comparat}. We also select only clusters with peak contamination fraction $f_{\text{count}} < 0.3$, which reduces the contamination level to $6\%$ \citep{2022Klein}. With these quality cuts, the number of clusters from the catalog reduces from $542$ to $477$. The complete description of eFEDS is available in \citet{2022Liu} and \cite{2022Klein}.

\subsubsection{XXL} \label{subsec:xxl}
The XXL project is a survey using \textit{XMM-Newton} to cover two $25$ deg$^2$ fields with observations as sensitive as a few times $10^{-15}$ ergs$^{-1}$ cm$^{-2}$ in the $0.5$ - $2$ keV band. The catalog consists of $365$ clusters at redshift $0$ to $1.2$. The redshifts comes from the spectroscopic follow-up campaigns as well as existing spectroscopic redshift data in the area. \cite{2016Pierre}, \cite{2016Pacaud}, and \cite{2018Adami} provide more information on this survey.

\subsubsection{CAMIRA} \label{subsec:camira}
The HSC CAMIRA cluster catalog identifies clusters in the HSC wide area using the CAMIRA algorithm \citep{2014Oguri}. The algorithm uses an SPS model from \cite{2003Bruzual} to fit the photometric data from the galaxies. The fit is used to calculate the likelihood of the galaxies being on the red-sequence as a function of redshift. Galaxies with spectroscopic data are used to calibrate the SPS model. A non-integer numeric parameter is defined for each galaxy as a function of redshift, such that the sum of the parameters is richness. Peaks inside a three-dimensional richness map are identified as cluster candidates. The richness is defined as the number of red galaxies with stellar mass $M_{\star}>10^{10.2}M_{\odot}$ in a circular aperture of radius $R<1\;h^{-1}$Mpc.  We compare our candidates to the cluster catalog derived from the HSC DR3 \citep{2022Aihara}. The catalog contains $7319$ clusters with $0.1<$ $z$ $<1.375$. The details of the algorithm and the HSC CAMIRA cluster catalog characterization are available in \cite{2014Oguri} and \cite{2018Oguri}. 

\subsubsection{redMaPPer} \label{subsec:redmapper}
The redMaPPer DR8 catalog is a cluster catalog generated using the red-sequence matched-filter Probabilistic Percolation cluster finder, redMaPPer \citep{2014Rykoff}, on SDSS DR8. The algorithm detects clusters using the assumption that galaxies in clusters are old and red galaxies. The overdensity of these passive galaxies is used to identify clusters. The DR8 catalog identifies $26,311$ cluster candidates in $0.08 <$ $z$ $< 0.6$. The full description of this catalog is in \cite{2016Rykoff}.

\subsubsection{MaDCoWS} \label{subsec:madcows1}
MaDCoWS \citep{2019Gonzalez} uses optical rejection and infrared color-selection to select galaxies at $z$ $>0.8$. Overdensities of these galaxies are identified as galaxy clusters. The smoothed density maps in MaDCoWS are derived from a Difference-of-Gaussian kernel similar to the one used in this work, though with fixed angular sizes of $38.2"$ and $3.82'$ for the inner and outer kernels, rather than the fixed physical kernel sizes given in \S 5.1. The \textit{WISE}-Pan-STARRS catalog to which we compare utilizes AllWISE W1 and W2 with Pan-STARRS $i$-band. \cite{2019Gonzalez} present the highest peak detections as cluster candidates, of which $2433$ are from the \textit{WISE}-Pan-STARRS region, and $250$ are from \textit{WISE}-SuperCOSMOS region. Of these, $1723$ of \textit{WISE}-Pan-STARRS cluster candidates have follow-up observations using $Spitzer$ which produce the photometric redshift and richness. The photometric redshifts of the cluster candidates are at $0.7<$ $z$ $<1.5$, and the $38$ clusters with spectroscopic redshifts show that the photometric redshifts have an uncertainty of $\sigma_z/(1+z)\simeq0.036$. CARMA observations indicate that the median mass for the WISE-Pan-STARRS clusters is $M_{500}=1.6^{+0.7}_{-0.8} \times 10^{14} M_{\odot}$ \citep{2019Gonzalez}.

\subsection{Cross-matching of Cluster Catalogs} \label{subsec:crossmatch cluster}
We employ a simple rank-order matching. Starting with the highest S/N candidates in our search, we match in descending S/N order. Cluster candidates from other surveys are considered matches if they lie within $1$ Mpc and $\delta z/(1+z_{\rm other})=0.2$ of a MaDCoWS2 candidate. The closest candidate within this projected radius is considered a match. We exclude clusters from other catalogs that have $f_{\rm area}>0.2$  (as defined in \S \ref{subsec:algorithm}) within $30^{\prime\prime}$ or $120^{\prime\prime}$, rather than $f_{\rm area}>0.4$ in \S \ref{subsec:algorithm}. The reason for the aggressive $f_{\rm area}$ requirement here is to avoid misleading results in assessing the completeness of the MaDCoWS2 catalog in our comparison. This cut ensures that the external clusters are not at the edge of the rejected area of our search.

\subsection{Fraction of Clusters Rediscovered} \label{subsec:rediscovered fraction}

\begin{deluxetable*}{cccccc}[htbp]
\tablecaption{Fit values from the error functions in Figure \ref{fig:completness} and the total rediscovered fractions. \label{tab:completeness}}
\tablehead{\colhead{Quantity} & \colhead{$\mu$} & \colhead{$\sigma$} & \colhead{$\mu $}& \colhead{$\sigma$} & \colhead{Total $f_{RD}$} \\ 
\colhead{} & \colhead{S/N $\geq5$} & \colhead{S/N $\geq5$} & \colhead{S/N $\geq7$} & \colhead{S/N $\geq7$} & \colhead{S/N $\geq5$} } 
\startdata
log$_{10}(M_{500}^{\rm ACT})$ & 13.91 & 0.44 & 14.17 & 0.37 & 0.92 \\
log$_{10}(L_{\rm 500kpc}^{\rm eFEDS})$ & 43.28 & 1.08 & 44.02 & 1.53 & 0.62 \\
log$_{10}(M_{\rm gas,500kpc}^{\rm XXL})$ & 12.44 & 0.47 & 12.79 & 0.62 & 0.66 \\
$\lambda_{\rm CAMIRA}$ & 17.98 & 15.36 & 31.63 & 16.68 & 0.60 \\
$\lambda_{\rm redMaPPer}$ & -0.01 & 44.01 & 33.80 & 32.85 & 0.87 \\
$\lambda_{\rm MaDCoWS}$ & 24.13 & 16.40 & 42.36 & 14.52 & 0.60 \\
\enddata
\tablecomments{The fit values of $\mu$ and $\sigma$ for the error functions in Figure \ref{fig:completness}. $\mu$ represents the position on the x-axis with $50\%$ completeness ($f_{RD}=0.5$), and $\sigma$ is the standard deviation of the error function. The total $f_{RD}$ is the fraction of clusters rediscovered from the total number of clusters. $M_{500}^{\text{ACT}}$ and $M_{gas,500kpc}^{\text{XXL}}$ are in $M_\odot$ while $L_{500kpc}^{\text{eFEDS}}$ is in units of $\mathrm{erg\;s}^{-1}$.}
\end{deluxetable*}

\begin{figure*}[htbp]
\centering
\includegraphics[width=0.9\textwidth]{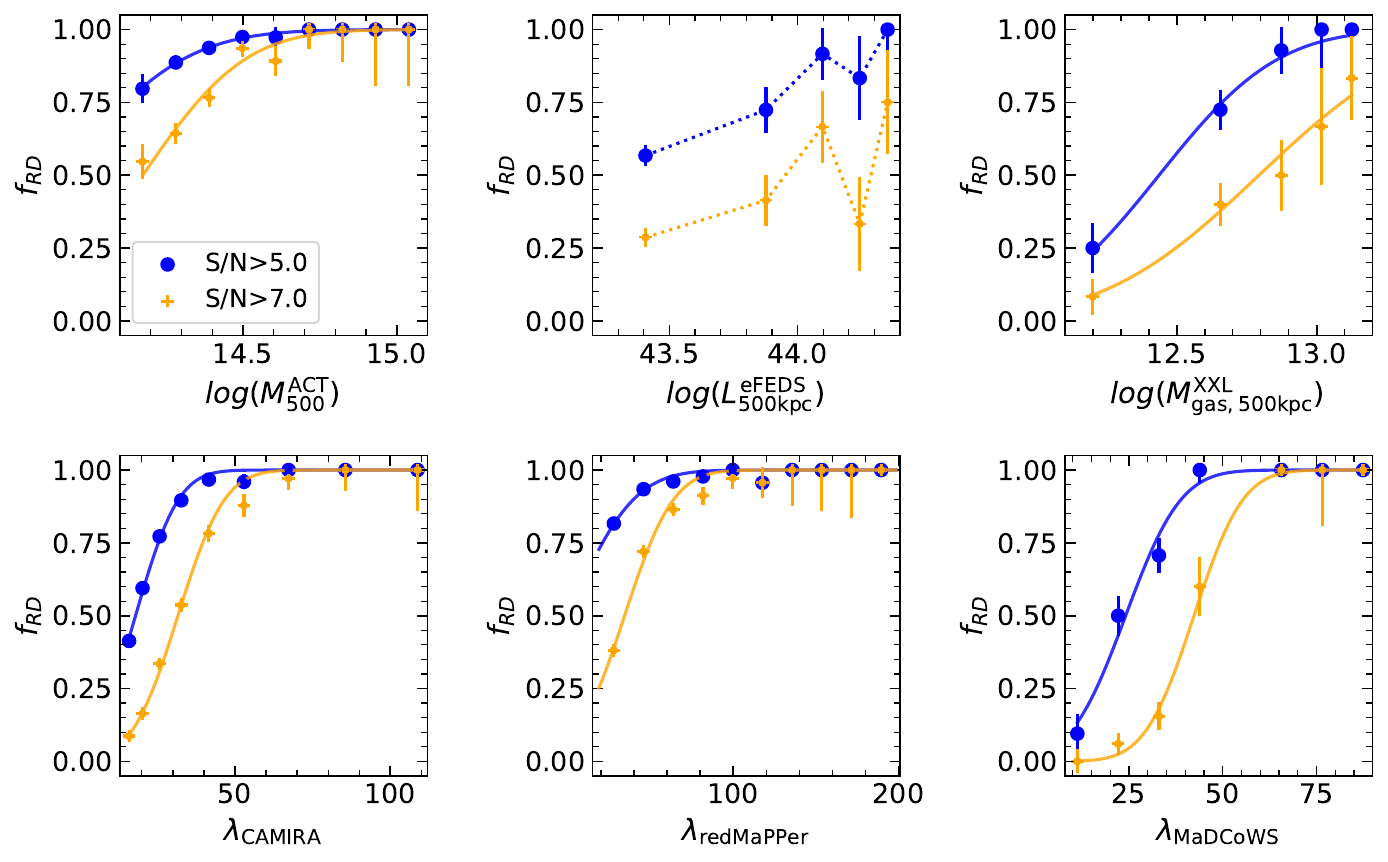}
\caption{The fraction of clusters rediscovered from external cluster catalogs as a function of cluster properties. For clarity, we omit the upper error bar when $f_{RD}=1$. Different line colors designate different S/N cutoffs for MaDCoWS2. The solid lines show the error function fitted to the data points. The dashed lines in the top center panel show data points that are not well fit by an error function. $M_{500}^{\rm ACT}$ and $M_{\rm gas,500kpc}^{\rm XXL}$ are in $M_\odot$, while $L_{\rm 500kpc}^{\rm eFEDS}$ is in units of $\mathrm{erg\;s}^{-1}$.}
\label{fig:completness}
\end{figure*}

\begin{figure*}[htbp]
\centering
\includegraphics[width=0.9\textwidth]{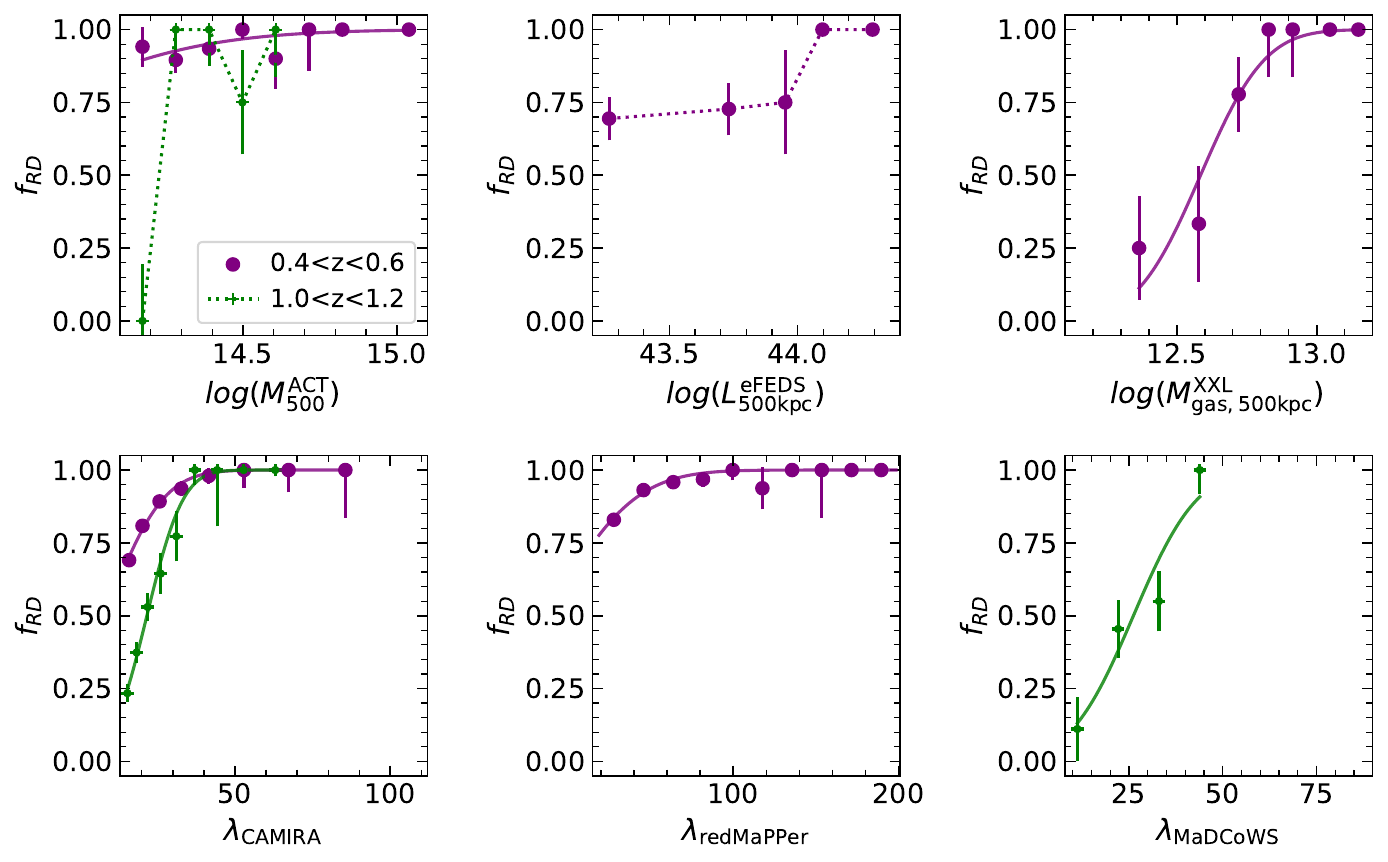}
\caption{The same as Figure \ref{fig:completness}, but different colors now show different redshift ranges for the external catalogs. Purple indicates $0.4<$ z$_{\rm ref}<0.6$, while green indicates $1<$ z$_{\rm ref}<1.2$. This figure adopt an S/N threshold of S/N $\geq5$.}
\label{fig:completness2}
\end{figure*}

In Figures \ref{fig:completness} and \ref{fig:completness2}, we investigate the fraction of galaxy clusters from external catalogs that we rediscover ($f_{RD}$) to assess the completeness of our search. We define $f_{RD}$ as the number of matches from external catalogs in a given bin divided by the total number of external clusters in that bin. The mass proxy we use to define the bins in each comparison survey are as follows: we use cluster mass $M_{500}^{\rm ACT}$ from ACT obtained by assuming universal pressure profile and \cite{2010Arnaud} scaling relation, luminosity $L_{\rm 500kpc}^{\rm eFEDS}$ from eFEDS, gas mass $M_{\rm gas,500kpc}^{\rm XXL}$ from XXL, and richness $\lambda$ from CAMIRA, redMaPPer, and MaDCoWS. The symbols in the Figures represent the rediscovery fraction within that bin. The lines represent the best fits of the data for an error function defined as
\begin{equation}
    erf(x) = \frac{1}{2}+\frac{1}{\sqrt{\pi}}\int_0^{x}e^{-t^2}dt.    
\end{equation}
where $x=(Q-\mu)/\sigma$ and $Q$ denotes of the parameters from the external catalogs used for binning. We perform the fit using the least squares function from the \texttt{lmfit} package \citep{2023Newville}.  Figure \ref{fig:completness} shows the results considering the clusters from the entire redshift range of the external catalog (with different plot colors indicating the S/N lower limit of our search). Figure \ref{fig:completness2} uses the S/N $\geq5$ limit, and shows two different redshift ranges ($0.4<z<0.6$ and $1<z<1.2$).

In the areas that overlap with our search, $553$, $236$, $87$, $5019$, $3071$, and $154$ clusters from ACT, eFEDS, XXL, CAMIRA, reMaPPer, and MaDCoWS exist. With the S/N $\geq5$ cutoff of our survey, the total rediscovered fraction for each external catalog is $92\%,\; 62\%,\; 66\%,\; 60\%,\; 87\%,$ and $60\%$, respectively. $10$, $3$, $1$, $45$, and $57$ clusters from ACT, eFEDS, XXL, CAMIRA, and reMaPPer respectively are not counted as matches because their MaDCoWS2 counterparts are rejected by $f_{\rm area}>0.4$.

Table \ref{tab:completeness} presents the fit values for $\mu$ and $\sigma$ which indicate the $50\%$ completeness and standard deviation of the error function. We omit the fit for eFEDS in Figure \ref{fig:completness} as the scatter between its observable and our S/N is high (see Figure \ref{fig:prop}), resulting in data that are not well fit by an error function. Figure \ref{fig:completness} illustrates that our search reaches $50\%$ completeness with $M_{500}^{\rm ACT}$ and $\lambda_{\rm redMapPPer}$ at values lower than the lower limit of $M_{500}^{\rm ACT}$ and $\lambda_{\rm redMapPPer}$. For S/N $\geq5$, the $f_{RD}$ values are $85\%$ for $M_{500}^{\rm ACT}=10^{14.23}\;M_\odot$, $M_{\rm gas,500kpc}^{\rm XXL}=10^{12.78}\;M_\odot$, $\lambda_{\rm CAMIRA}=29.25$, $\lambda_{\rm redMapPPer}=32.25$, and $\lambda_{\rm MaDCoWS}=36.15$. 

At $0.4<$ z$_{\rm ref}<0.6$, we find $94\%$, $72\%$, $71\%$, $81\%$, and $91\%$ of clusters from ACT, eFEDS, XXL, CAMIRA, and redMaPPer, respectively. At $1<$ z$_{\rm ref}<1.2$, we find $88\%$, $41\%$, and $52\%$ of clusters from ACT, CAMIRA, and MaDCoWS, respectively.

Figures \ref{fig:completness} and \ref{fig:completness2} suggest that MaDCoWS2 is effective in rediscovering most of the clusters from the others surveys shown in Figure \ref{fig:zdist2}. The highest $f_{RD}$ (0.92) is found for ACT, the only SZ-based comparison sample, which is expected given its comparatively high mass threshold. redMaPPer also has a high $f_{RD}$ (0.87). redMapper has a lower projected number density than MaDCoWS at all redshifts, and thus a high $f_{RD}$ is again expected.

For a more general comparison, one can consider the redshift range $0.4<z_{ref}<0.6$, for which the surface density of MaDCoWS2 clusters is higher than all of the surveys but XXL. We interpret the high $f_{RD}$ values at this epoch as an indication that our completeness is generally high at low redshift when the projected density of MaDCoWS2 is comparable to or greater than the comparison sample.  

Considering the full redshift range, the other two optical surveys have significantly lower $f_{RD}$ than redMapper. For CAMIRA, we interpret this as a consequence of the CAMIRA catalog having a comparable or higher projected number density of clusters at both low redshift and $z\sim1$, and hence the shallower depth of MaDCoWS2 is the driving factor.

Conversely, the projected number density of MaDCoWS clusters is quite low. While it is tempting to expect a high $f_{RD}$ for MaDCoWS, the MaDCoWS2 search is significantly different from that of MaDCoWS. MaDCoWS detected clusters using a simple color selection of galaxies at $z>0.7$, while
MaDCoWS2 detect clusters in a density cube created from photometric-redshifts based upon deeper multiband data. This leads to significant scatter in the MaDCoWS S/N compared to MaDCoWS2, primarily due to larger line-of-sight projections effects for MaDCoWS.

Finally, for the X-ray surveys there are two factors depressing the $f_{RD}$. First, at the lowest redshifts the surface densities of XXL and eFEDS significantly exceed that of MaDCoWS2. Second, the scatter evident between the X-ray and MaDCoWS2 observables in Figure \ref{fig:completness2} is an additional factor likely depressing the $f_{RD}$ values for these samples. This scatter is furthered discussed in \S \ref{properties}.

In addition to the catalogs used in  Figure \ref{fig:completness} and \ref{fig:completness2}, we check whether we rediscover three spectroscopically confirmed $z$ $>1.5$ clusters. They are JKCS 041, ClG J0218.3-0510, and SpARCS J022426-032330, from \citet[$z=1.803$]{2009Andreon}, \citet[$z=1.62$]{2010Papovich}, and \citet[$z=1.633$]{2013Muzzin}, respectively. We rediscover JKCS 041 and ClG J0218.3-0510, with S/N $=5.27$ and $5.14$, and z$_{\rm phot}$ $=1.830\pm0.082$ and $1.590\pm0.075$, respectively. We find an S/N $\sim 4.5$ overdensity of galaxies at the coordinate and redshift range of SpARCS J022426-032330, though it does not meet our S/N $>5$ detection criterion.  While this is not a large sample size, these are the only $1.5<z<2$ spectroscopically confirmed clusters from external catalogs in our unmasked survey area to which we can compare. The fact that two of them are also in MaDCoWS2 confirms the ability of MaDCoWS2 to find massive galaxy clusters at this redshift range.

\subsection{Cluster Properties} \label{properties}
\begin{figure*}[htbp]
\centering
\includegraphics[width=0.9\textwidth]{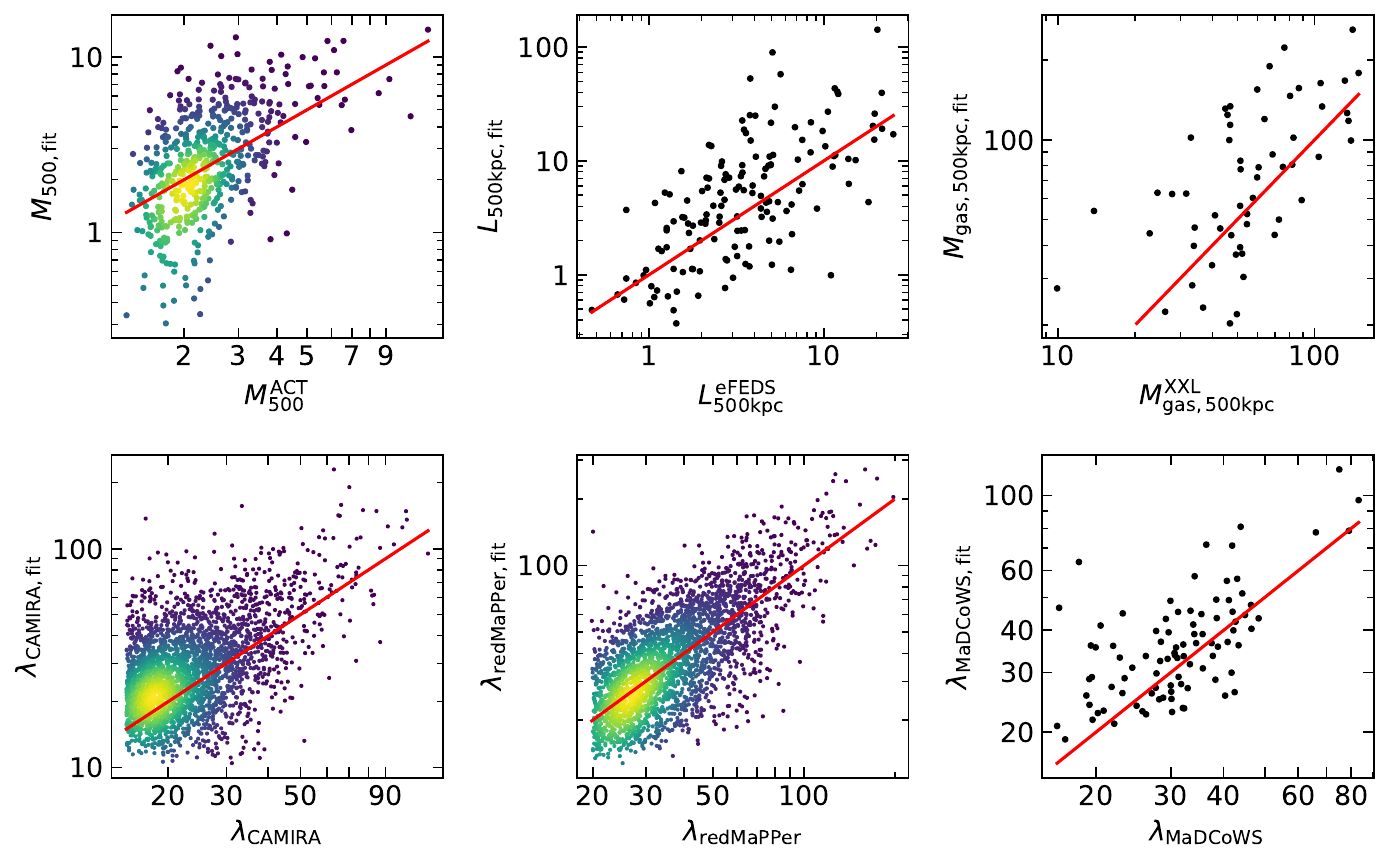}
\caption{Predicted values of cluster observables derived from MaDCOWS2 S/N-observable relations versus the published values of these quantities from external catalogs. These predicted values are derived using Equation \ref{powerlaw} with the best fit values in Table \ref{tab:propfit}. The red lines are the one-to-one lines for each quantity. $M_{500}^{\rm ACT}$ and $M_{\rm gas,500kpc}^{\rm XXL}$ are given in units of 10$^{14} M_\odot$ and 10$^{11} M_\odot$, respectively, while $L_{\rm 500kpc}^{\rm eFEDS}$ is given in units of $10^{43}$ erg s$^{-1}$.}
\label{fig:prop}
\end{figure*}

In Figure \ref{fig:prop}, we compare between the MaDCoWS2 S/N and cluster properties of external catalogs. We fit a power law function to the plot of our S/N versus these quantities using truncated linear regression via Markov chain Monte Carlo (MCMC) method from the \texttt{PyMC} package \citep{pymc2023}. The power law follows the equation
\begin{equation}
    \log_{10}\text{S/N} = a + b \log_{10}\text{Q} + c \log_{10}(1+z_{\rm ref})
\label{powerlaw}
\end{equation}
where $Q$ is any of the quantities from the external catalogs.  We report the fit values in Table \ref{tab:propfit}. We assess the scatter of the residual using the standard deviation of the residual $\sigma_{\text{res}}$ where residual$=\text{S/N}-\text{S/N}_\text{fit}$. Figure \ref{fig:prop} shows the cluster properties derived from our S/N versus the original cluster properties from other catalogs. The y-axis of each panel is derived by calculating $Q$ in equation \ref{powerlaw} using the fit results. 

The S/N of MaDCoWS2 has the lowest scatter in relation to $\lambda_{\text{MaDCoWS}}$ with $\sigma_{\text{res}}=0.99$, but the parameters are not as well constrained compared to those from other optical catalogs and ACT due to its lower number of data points. The scatter become larger for other comparisons, with $\sigma_{\rm res}=1.47,\;1.56,\;1.83,\;1.85$ and$,\;2.16$ for $\lambda_{\rm CAMIRA}$, $\lambda_{\rm redMaPPer}$, $M_{\rm gas,500kpc}^{\rm XXL}$, $L_{\rm 500kpc}^{\rm eFEDS}$, and $M_{500}^{\rm ACT}$, respectively.

It is interesting to note that the galaxy-based cluster catalogs generally have lower scatter in the relations between their observables and MaDCoWS2 SNR compared to ICM-based catalogs. We interpret this to be a reflection of intrinsic scatter which arises due to the two different techniques. If we consider the excess scatter beyond what is seen for CAMIRA and redMaPPer indicative of the physical scatter between the ICM and galaxy properties of clusters, this implies a physical scatter of $\sigma_{\rm phys}\simeq 1-1.1$ for optical versus X-ray surveys and $\sigma_{\rm phys}\simeq 1.5-1.6$ for optical versus SZ surveys due to the different detection methods.

\begin{deluxetable*}{cccccc}
\tablecaption{Fit values for power law fit to S/N versus external catalogs quantities. \label{tab:propfit}}
\tablehead{\colhead{Quantity} & \colhead{N(S/N$>5$)} & \colhead{$a$} & \colhead{{$b$}} & \colhead{$c$} & \colhead{$\sigma_{\rm res}$}}
\startdata
$M_{500}^{\rm ACT}$ & 506 & -4.3$\pm$0.4 & 0.381$\pm$0.03 & -0.998$\pm$0.08 & 2.16 \\
$L_{500kpc}^{\rm eFEDS}$ & 146 & -9.4$\pm$0.9 & 0.24$\pm$0.02 & -1.24$\pm$0.23 & 1.85 \\
$M_{gas,500kpc}^{\rm XXL}$ & 57 & -4.78$\pm$0.85 & 0.45$\pm$0.07 & -0.99$\pm$0.37 & 1.83 \\
$\lambda_{\rm CAMIRA}$ & 3036 & 0.35$\pm$0.03 & 0.43$\pm$0.02 & -0.62$\pm$0.06 & 1.47 \\
$\lambda_{\rm redMaPPer}$ & 2681 & 0.19$\pm$0.02 & 0.50$\pm$0.01 & -0.70$\pm$0.07 & 1.56 \\
$\lambda_{\rm MaDCoWS}$ & 92 & 0.20$\pm$0.16 & 0.54$\pm$0.09 & -0.71$\pm$0.32 & 0.99 \\
\enddata
\tablecomments{The fit values $a$, $b$, and $c$ for a power law fit between the S/N of this work and cluster quantities from other surveys (equation \ref{powerlaw} in \S \ref{properties}). N represents the number of clusters available in each fit. $\sigma_{\text{res}}$ represents the standard deviation of the residual where residual$=\text{S/N}-\text{S/N}_\text{fit}$.}
\end{deluxetable*}

\subsection{Redshift Comparison} \label{subsec:Redshift comparison}
\begin{figure}[htbp]
\centering
\includegraphics[width=0.9\columnwidth]{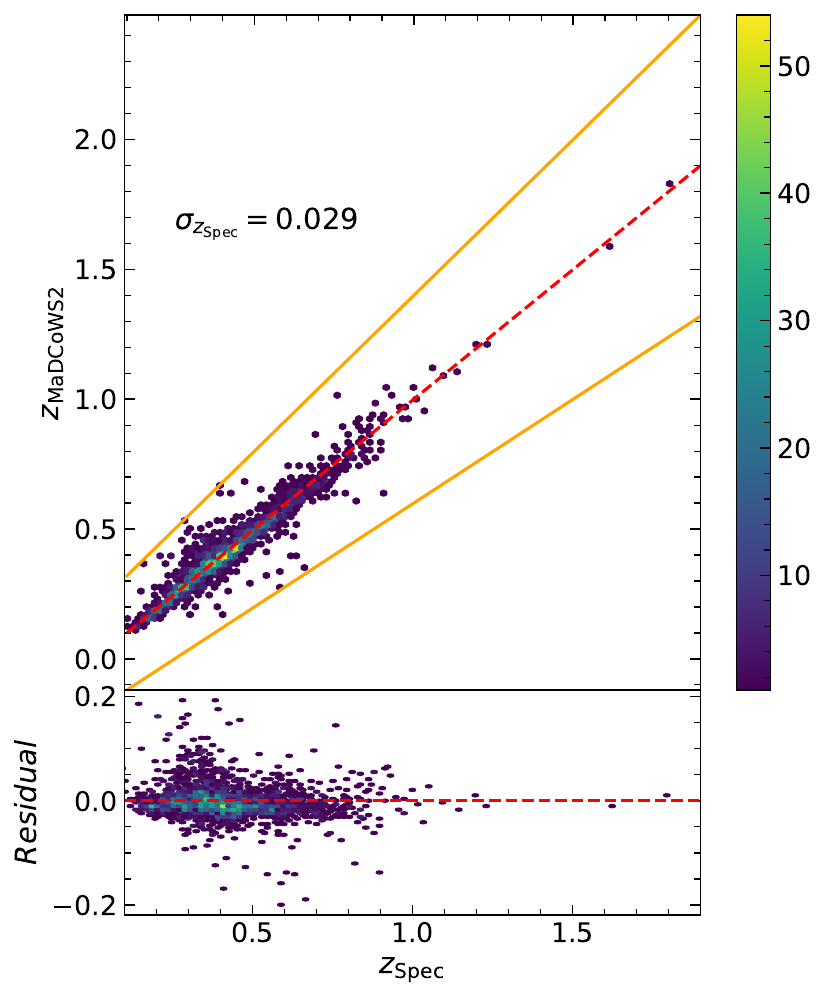}
\caption{Comparison between the photometric redshifts of galaxy cluster candidates in this work and spectroscopic redshifts from external catalogs. The red dashed line indicates a 1-to-1 line in the top panel and zero residual in the bottom panel. The bottom panel displays the residual $\delta z /(1+z_{\rm Spec})$ as a function of spectroscopic redshift. The colors indicate the density of points on the plot.}
\label{fig:zcompspec}
\end{figure}

 We compare the photometric redshift from our survey to those from external catalogs. The clusters shown in Figure \ref{fig:zcompspec} are considered matches in our cluster catalog crossmatching process. This means they all have $\delta z/(1+z_{\rm other})<0.2$. The medians $\Tilde{\delta z}$, mean $\overline{\delta z}$, and standard deviations $\sigma_{z}$ of $\delta z/(1+z_{\rm other})$ of all the comparisons are presented in Table \ref{tab:zcomptable}. For each external catalog, we calculated $\Tilde{\delta z}$, $\overline{\delta z}$, and $\sigma_{z}$ using all redshifts available in that catalog. The $\Tilde{\delta z}_{\rm Spec}$, $\overline{\delta z}_{\rm Spec}$, and $\sigma_{z_{\rm Spec}}$ are calculated using only the spectroscopic redshifts. The photometric redshifts from MaDCoWS2 agree with the redshifts available in other searches as $\Tilde{\delta z}<0.02$ and $\sigma_{z}<0.06$ for all of the comparisons. Comparing to the spectroscopic redshifts from the external catalogs, $\Tilde{\delta z}_{\rm Spec} = -0.002$ and $\sigma_{z_{\rm Spec}} = 0.029$, illustrating the accuracy of the photometric redshift of our candidates compared to the spectroscopic redshifts (Figure \ref{fig:zcompspec}).

\begin{deluxetable}{ccccccc}[htbp]
\tablecaption{The median, mean, and standard deviation of $\delta z/(1+z_{\rm other})$ for cluster candidates at S/N $\geq5$ \label{tab:zcomptable}}
\tablehead{\colhead{Survey} & \colhead{$\Tilde{\delta z}$} & \colhead{$\overline{\delta z}$} & \colhead{$\sigma_{z}$} & \colhead{$\Tilde{\delta z}_{\rm Spec}$} & \colhead{$\overline{\delta z}_{\rm Spec}$} & \colhead{$\sigma_{z_{\rm Spec}}$}} 

\startdata
ACT & -0.007 & -0.002 & 0.03 & -0.01 & -0.01 & 0.018 \\
eFEDS & -0.001 & 0.01 & 0.038 & 0.003 & 0.003 & 0.036 \\
XXL & \nodata & \nodata & \nodata & -0.013 & -0.01 & 0.025 \\
CAMIRA & -0.004 & 0.0001 & 0.036 & -0.006 & -0.006 & 0.028 \\
redMaPPer & -0.005 & 0 & 0.029 & -0.005 & -0.005 & 0.029 \\
MC1 & 0.005 & 0.007 & 0.056 & -0.009 & -0.009 & 0.029 \\
% &  \\
\enddata
\tablecomments{The quantities without the $\rm Spec$ subscript are calculated using both photometric and spectroscopic redshifts from that external catalog. The quantities with the $\rm Spec$ subscript are calculated using only the spectroscopic redshift.}
\end{deluxetable}

\subsection{Offset of the Position of Detections} \label{subsec:offset}
\begin{figure*}[htbp]
\centering
\includegraphics[width=0.9\textwidth]{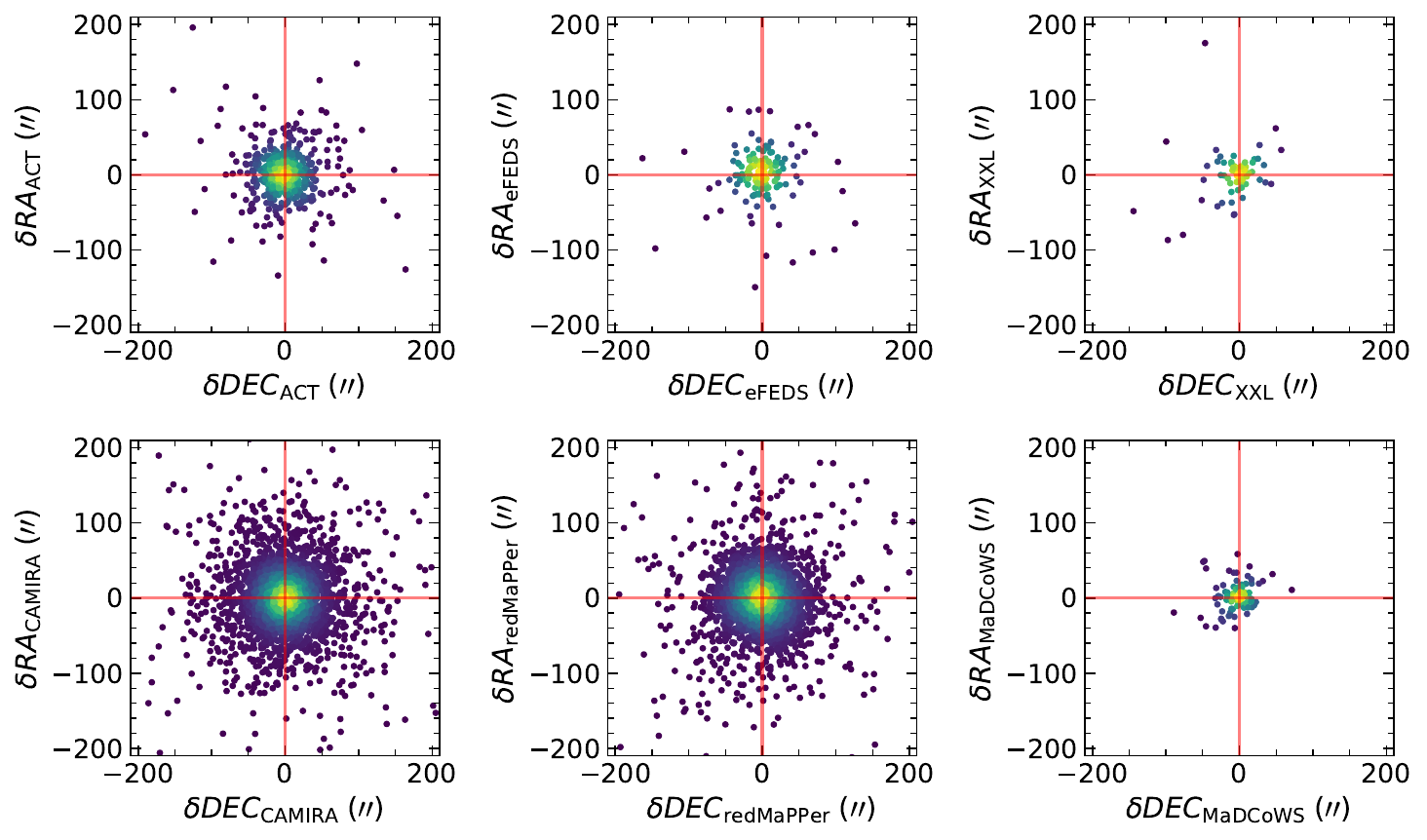}
\includegraphics[width=0.9\textwidth]{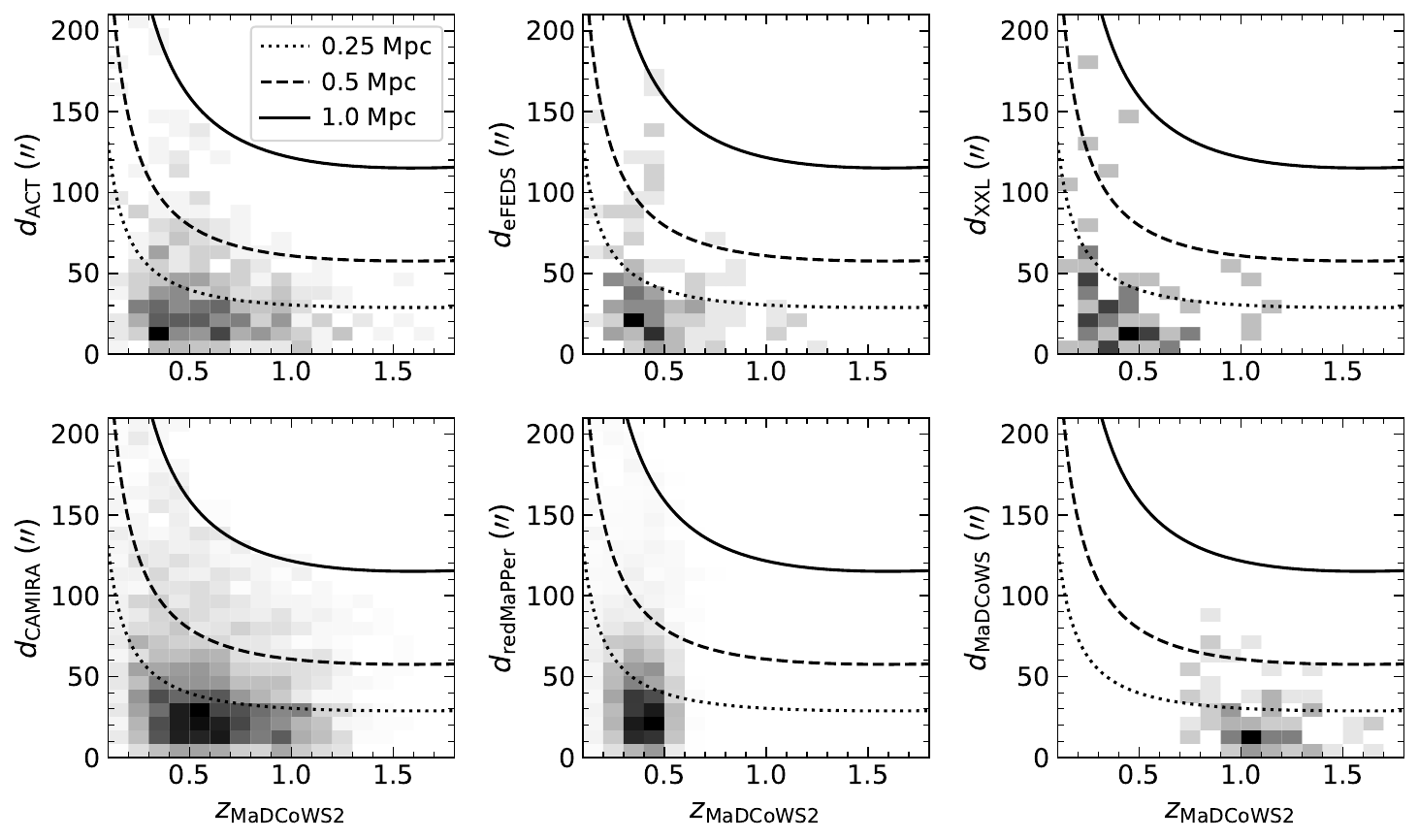}
\caption{Positional offset comparison with external catalogs, where color indicates the density of points. The bottom two rows show the radial offset $d$ as a function of photometric redshift. The lines denote constant physical radii, as indicated.} 
\label{fig:offset}
\end{figure*}

\begin{figure}[htbp]
\centering
\includegraphics[width=0.9\columnwidth]{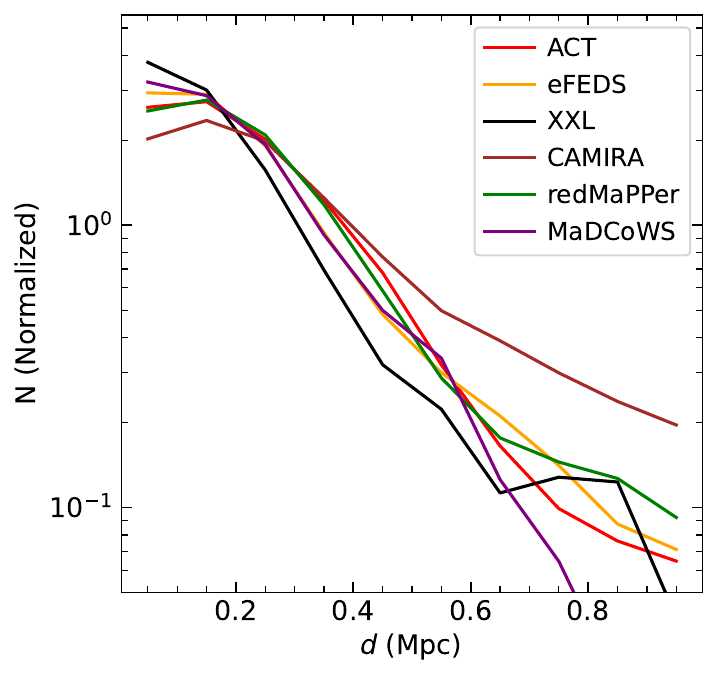}
\caption{Radial offset smoothed distribution from external catalogs. The distribution is smoothed by a Gaussian filter with $0.3$ Mpc radius and standard deviation of $0.75$}
\label{fig:offset2}
\end{figure}

Figure \ref{fig:offset} and \ref{fig:offset2} show the offset between the positions of clusters from the external catalogs compared to the MaDCoWS2 positions. The medians offsets are $1.99",\;3.21",\;3.71",\;0.5",\;0.47",$ and $2.8"$ for ACT, eFEDS, XXL, CAMIRA, reMaPPer, and MaDCoWS, respectively. The standard deviations are $35",\;43",\;34",\;47",\;44",$ and $21"$, respectively. The MaDCoWS2 offsets concentrate below the $0.25$ Mpc for all surveys we considered. There is no systematic offset between MaDCoWS2 cluster candidate positions and those from external catalogs.

\section{Summary} \label{sec:summary}
MaDCoWS2 Data Release 1 provides a catalog of $22,970$ galaxy cluster candidates with S/N $>5$ at $0.1<z<2$. These clusters are detected using a CatWISE2020-selected galaxies. The survey provides $1312$ candidates at $z>1.5$. The photometric redshift with full PDF in this search is derived from $g$, $r$, $z$, W1, and W2 data from DECaLS from LS DR9 and CatWISE2020. The search identifies the overdensity of galaxies in three dimensions (sky positions and photometric redshifts) using the PZWav algorithm. The area of the search is $1461$ deg$^2$ ($1838$ deg$^2$ without masking), covering the equatorial fields of HSC DR3.

The wide area of this catalog overlaps with catalogs from ACT \citep{2021Hilton}, eFEDS \citep{2022Liu}, XXL \citep{2018Adami}, CAMIRA \citep{2018Oguri}, redMaPPer \citep{2016Rykoff}, and MaDCoWS \citep{2019Gonzalez}. For a S/N $>5$ cutoff, we detect $92\%,\; 62\%,\; 66\%,\; 60\%,\; 87\%,$ and $60\%$, respectively, of the clusters in the surveys. For all the surveys compared, the median and standard deviation of $\delta z/(1+z_{\rm Ref})$ are below $0.02$ and $0.06$. The median positional offsets are under $0.25$ Mpc. We also investigate the relation between our S/N and the cluster mass $M_{\rm 500}^{\rm ACT}$ from ACT (calibrated by \citet{2010Arnaud} scaling relation), the luminosity $L_{\rm 500kpc}^{\rm eFEDS}$ from eFEDS, the gas mass $M_{\rm gas,500kpc}^{\rm XXL}$ from XXL, and the richnesses $\lambda$ from CAMIRA, redMaPPer, and MaDCoWS by fitting power law functions. The relation with $\lambda_{\rm MaDCoWS}$ provides the lowest scatter, followed by $\lambda_{\rm CAMIRA}$ and $\lambda_{\rm redMaPPer}$.  

In the future, the full data of DECaLS will be included in the survey, expanding the survey coverage area of MaDCoWS2 to $>8000$ deg$^2$ with an effective area $>6000$ deg$^2$. This will allow a more rigorous assessment of the quality of MaDCoWS2 since the sample sizes of the external catalogs will be larger. Future cluster surveys such as those from $Euclid$ and LSST will benefit from the full search that will be available from the next release of this survey. The full search will not only provide cluster candidates to these upcoming surveys can utilize, but also enable exploration of the systematics from a wide optical/IR survey.

\acknowledgments
We thank Edward Schlafly for his insights into the DESI Legacy Survey data and Jean-Baptiste Melin for his development of the S/N definition employed for MaDCoWS2. We appreciate comments from the anonymous referee that improved the manuscript.

The authors acknowledge University of Florida Research Computing for providing computational resources and support that have contributed to the research results reported in this publication.

This material is based upon work supported by the National Science Foundation under Grant No. 2108367. The work of T.C., P.E., and D.S. was carried out at the Jet Propulsion Laboratory, California Institute of Technology, under a contract with the National Aeronautics and Space Administration (80NM0018D0004).

This publication makes use of data products from the \textit{Wide-field Infrared Survey Explorer}, which is a joint project of the University of California, Los Angeles, and the Jet Propulsion Laboratory/California Institute of Technology, funded by the National Aeronautics and Space Administration.

This research has made use of the NASA/IPAC Infrared Science Archive, which is funded by the National Aeronautics and Space Administration and operated by the California Institute of Technology.

This research is based upon the Dark Energy Camera Legacy Survey (DECaLS; PIs: David Schlegel and Arjun Dey). The Legacy Surveys imaging of the DESI footprint is supported by the Director, Office of Science, Office of High Energy Physics of the U.S. Department of Energy under Contract No. DE-AC02-05CH1123, by the National Energy Research Scientific Computing Center, a DOE Office of Science User Facility under the same contract; and by the U.S. National Science Foundation, Division of Astronomical Sciences under Contract No.AST-0950952 to NOAO.

This project used data obtained with the Dark Energy Camera (DECam), which was constructed by the Dark Energy Survey (DES) collaboration. Funding for the DES Projects has been provided by the U.S. Department of Energy, the U.S. National Science Foundation, the Ministry of Science and Education of Spain, the Science and Technology Facilities Council of the United Kingdom, the Higher Education Funding Council for England, the National Center for Supercomputing Applications at the University of Illinois at Urbana-Champaign, the Kavli Institute of Cosmological Physics at the University of Chicago, Center for Cosmology and Astro-Particle Physics at the Ohio State University, the Mitchell Institute for Fundamental Physics and Astronomy at Texas A$\&$M University, Financiadora de Estudos e Projetos, Fundação Carlos Chagas Filho de Amparo, Financiadora de Estudos e Projetos, Fundação Carlos Chagas Filho de Amparo à Pesquisa do Estado do Rio de Janeiro, Conselho Nacional de Desenvolvimento Científico e Tecnológico and the Ministério da Ciência, Tecnologia e Inovação, the Deutsche Forschungsgemeinschaft and the Collaborating Institutions in the Dark Energy Survey. The Collaborating Institutions are Argonne National Laboratory, the University of California at Santa Cruz, the University of Cambridge, Centro de Investigaciones Enérgeticas, Medioambientales y Tecnológicas–Madrid, the University of Chicago, University College London, the DES-Brazil Consortium, the University of Edinburgh, the Eidgenössische Technische Hochschule (ETH) Zürich, Fermi National Accelerator Laboratory, the University of Illinois at Urbana-Champaign, the Institut de Ciències de l'Espai (IEEC/CSIC), the Institut de Física d'Altes Energies, Lawrence Berkeley National Laboratory, the Ludwig-Maximilians Universität München and the associated Excellence Cluster Universe, the University of Michigan, the National Optical Astronomy Observatory, the University of Nottingham, the Ohio State University, the University of Pennsylvania, the University of Portsmouth, SLAC National Accelerator Laboratory, Stanford University, the University of Sussex, and Texas A$\&$M University.

The Photometric Redshifts for the Legacy Surveys (PRLS) catalog used in this paper was produced thanks to funding from the U.S. Department of Energy Office of Science, Office of High Energy Physics via grant DE-SC0007914.

The Siena Galaxy Atlas was made possible by funding support from the U.S. Department of Energy, Office of Science, Office of High Energy Physics under Award Number DE-SC0020086 and from the National Science Foundation under grant AST-1616414.

\vspace{5mm}
\facility{IRSA,WISE}

\software{PZWav \citep{2014Gonzalez}, Astropy \citep{astropy:2013, astropy:2018, astropy:2022}, Matplotlib \citep{Hunter:2007}, NumPy \citep{harris2020array}, pandas \citep{the_pandas_development_team_2023_10107975,mckinney-proc-scipy-2010}, lmfit \citep{newville_matthew_2014_11813}, SciPy \citep{2020Virtanen}, PyMC \citep{pymc2023}}

\newpage
\clearpage

\appendix
\section{Redshift comparison}
\begin{figure*}[htbp]
\centering
\includegraphics[width=0.9\textwidth]{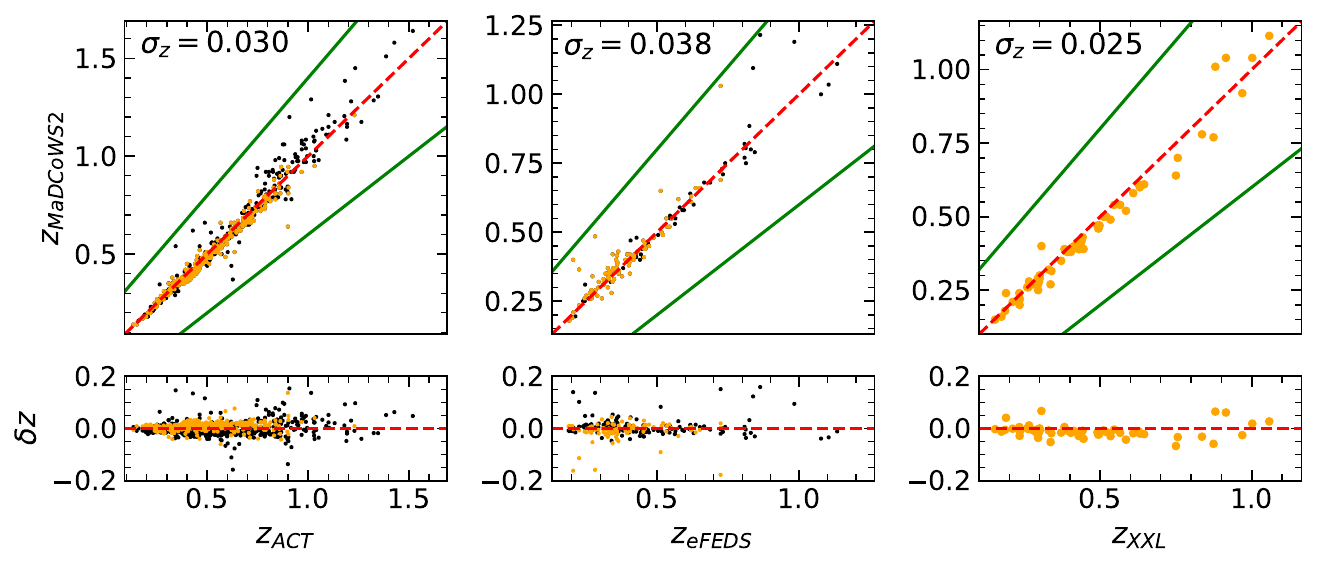}
\includegraphics[width=0.9\textwidth]{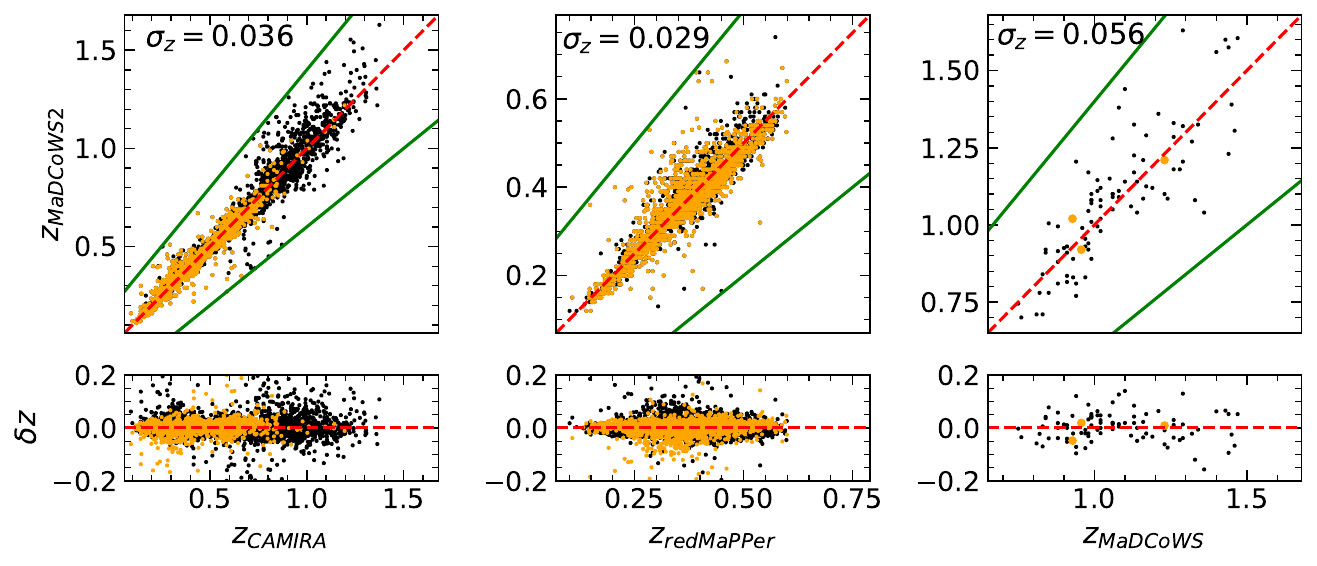}
\caption{Comparison between redshifts derived in this work against those from external catalogs, including both photometric (black) and spectroscopic (orange) redshifts.  The red dashed lines indicate the 1-to-1 line. The green lines indicate the residual $\delta z/(1+z_{\rm Ref})=\pm0.2$. $\sigma_z$ is the standard deviation of $\delta z/(1+z_{\rm Ref})$ considering both photometric and spectroscopic redshifts.} 
\label{fig:zcomp}
\end{figure*}

Figure \ref{fig:zcomp} compares photometric redshifts from the MaDCoWS2 catalog to the photometric (black) and spectroscopic (orange) redshifts from external catalogs. The statistics for these plots are presented in Table \ref{tab:zcomptable}.

\section{Catalog Table}
The 10 highest S/N MaDCoWS2 clusters are presented here. The full table of 22,970 clusters is available electronically and included with this paper as a machine-readable table.
\begin{rotatetable*}
\begin{deluxetable*}{lllccccclccc}
\tabletypesize{\footnotesize}
\tablecaption{MaDCoWS2 galaxy cluster candidates, in S/N order. \label{tab:sample2}}
\tablehead{\colhead{Name} & \colhead{RA} & \colhead{DEC} & \colhead{$z_{\rm phot}$} & \colhead{$\epsilon_{z_{\rm phot}}$} & \colhead{S/N} & \colhead{S/N$_G$} & \colhead{Name$_{\rm fg}$} & \colhead{Literature Name} & \colhead{LitRef} & \colhead{zspec} & \colhead{zspecRef} \\ 
\colhead{} & \colhead{Deg} & \colhead{Deg} & \colhead{} & \colhead{} & \colhead{} & \colhead{} & \colhead{} & \colhead{} & \colhead{} & \colhead{} & \colhead{}} 
\startdata
MOO2 J02199+01304 &  34.976 &   1.507 & 0.350 & 0.039 & 20.2 & 30.6 & \nodata & ACT-CL J0219.9+0130 & [2,12,14] & 0.365 & 2 \\
MOO2 J22117-03491 & 332.945 &  -3.819 & 0.410 & 0.041 & 20.1 & 31.0 & \nodata & ACT-CL J2211.7-0349 & [2,3,12] & \nodata & \nodata \\
MOO2 J23083-02114 &  347.084 &  -2.190 & 0.310 & 0.038 & 20.1 & 30.5 & \nodata & ACT-CL J2308.3-0211 & [2,3,12] & 0.289 & 2 \\
MOO2 J02398-01345 &   39.972 &  -1.576 & 0.350 & 0.039 & 19.7 & 29.6 & \nodata & Abell  370 & [1,2,3,12] & 0.373 & 1\\
MOO2 J01400-05550 &   25.004 &  -5.917 & 0.410 & 0.041 & 18.7 & 28.4 & \nodata & ACT-CL J0140.0-0554 & [2,5,12] & 0.450 & 2 \\
MOO2 J09352+00476 &  143.803 &   0.793 & 0.360 & 0.039 & 18.3 & 27.0 & \nodata & ACT-CL J0935.2+0048 & [2,3,10,12,14] & 0.355 & 2\\
MOO2 J09161-00231 &  139.026 &  -0.386 & 0.300 & 0.038 & 18.1 & 26.6 & \nodata & ACT-CL J0916.1-0024 & [2,3,5,12] & 0.318 & 2\\
MOO2 J09566-02235 &  149.160 &  -2.392 & 0.490 & 0.043 & 17.0 & 25.2 & \nodata & ACT-CL J0956.6-0223 & [2] & 0.497 & 2\\
MOO2 J01318-13370 &   22.951 & -13.618 & 0.200 & 0.035 & 16.9 & 22.0 & \nodata & Abell  209 & [1,2,3] & 0.206 & 1\\
MOO2 J01004+04484 &   15.118 &   4.808 & 0.350 & 0.039 & 16.7 & 24.2 & \nodata & RMJ010032.8+044900.5 & [12] & 0.342 & 12\\
\enddata
\tablecomments{Example lines from the full MaDCoWS2 catalog, showing the 10 highest S/N candidates. The full table is provided electronically in machine-readable form. Reference: [1] \cite{1989Abell}, [2] \cite{2021Hilton}, [3] \cite{2016Planck}, [4] \cite{1990Gioia}, [5] \cite{2001Ebeling}, [6]\cite{2012Muzzin,2017Balogh}, [7] \cite{2009Andreon}, [8] \cite{2010Papovich}, [9] \cite{2018Adami}, [10] \cite{2022Liu}, [11] \cite{2012Mehrtens}, [12] \cite{2016Rykoff}, [13] \cite{2019Gonzalez}, [14] \cite{2018Oguri}, [15] \cite{2017Radovich}.}
\end{deluxetable*}
\end{rotatetable*}

\newpage
\clearpage

\bibliography{MaDCoWS2DR1}{}
\bibliographystyle{yahapj}

\end{document}